\documentclass{aastex63}

\hypersetup{linkcolor=red,citecolor=cyan,filecolor=green,urlcolor=magenta}

\received{}
\revised{}
\accepted{}
\shorttitle{Parameters estimation}
\shortauthors{Pei et al.}

\begin{document}

\title{The Estimation of Fundamental Physics Parameters for {\it Fermi}-LAT Blazars}

\correspondingauthor{Zhiyuan Pei}
\email{peizy@gzhu.edu.cn}
\correspondingauthor{Junhui Fan}
\email{fjh@gzhu.edu.cn}

\author[0000-0002-4970-3108]{Zhiyuan Pei}
\affiliation{Center for Astrophysics, Guangzhou University, Guangzhou 510006, People's Republic of China}
\affiliation{Astronomy Science and Technology Research Laboratory of Department of Education of Guangdong Province, Guangzhou 510006, People's Republic of China}
\affiliation{Key Laboratory for Astronomical Observation and Technology of Guangzhou, Guangzhou 510006, People's Republic of China}

\author[0000-0002-5929-0968]{Junhui Fan}
\affiliation{Center for Astrophysics, Guangzhou University, Guangzhou 510006, People's Republic of China}
\affiliation{Astronomy Science and Technology Research Laboratory of Department of Education of Guangdong Province, Guangzhou 510006, People's Republic of China}
\affiliation{Key Laboratory for Astronomical Observation and Technology of Guangzhou, Guangzhou 510006, People's Republic of China}

\author{Jianghe Yang}
\affiliation{Department of Physics and Electronics Science, Hunan University of Arts and Science, Changde 415000, China}

\author{Danyi Huang}
\affiliation{Center for Astrophysics, Guangzhou University, Guangzhou 510006, People's Republic of China}
\affiliation{Astronomy Science and Technology Research Laboratory of Department of Education of Guangdong Province, Guangzhou 510006, People's Republic of China}
\affiliation{Key Laboratory for Astronomical Observation and Technology of Guangzhou, Guangzhou 510006, People's Republic of China}

\author{Ziyan Li}
\affiliation{Center for Astrophysics, Guangzhou University, Guangzhou 510006, People's Republic of China}
\affiliation{Astronomy Science and Technology Research Laboratory of Department of Education of Guangdong Province, Guangzhou 510006, People's Republic of China}
\affiliation{Key Laboratory for Astronomical Observation and Technology of Guangzhou, Guangzhou 510006, People's Republic of China}



\begin{abstract}

Aiming to delineate the physical framework of blazars, we present an effective method to estimate four important parameters based on the idea proposed by \citet{BK95}, including the upper limit of central black hole mass $M$, the Doppler factor $\delta$, the distance along the axis to the site of the $\gamma$-ray production $d$ (which then can be transformed into the location of $\gamma$-ray-emitting region $R_{\gamma}$) and the propagation angle with respect to the axis of the accretion disk $\Phi$. To do so, we adopt an identical sample with 809 {\it Fermi}-LAT-detected blazars which had been compiled in \citet{Pei20PASA}. These four derived parameters stepping onto the stage may shed new light on our knowledge regarding $\gamma$-ray blazars. With regard to the paper of \citet{BK95}, we obtain several new perspectives, mainly in: (1) putting forward an updated demarcation between BL Lacs and FSRQs based on the relation between broad-line region luminosity and disk luminosity both measured in Eddington units, i.e., $L_{\rm disk}/L_{\rm Edd}=4.68\times10^{-3}$, indicating that there are some differences between BL Lacs and FSRQs on the accretion power in the disk; (2) proposing that there is a so-called `appareling zone', a potential transition field between BL Lacs and FSRQs where the changing-look blazars perhaps reside; (3) the location of $\gamma$-ray emission region is principally constrained outside the broad-line region, and for some BL Lacs are also away from the dusty molecular torus, which means the importance of emission components in the jet.      

\end{abstract}



\section{INTRODUCTION}

Blazars are a particular class of radio-loud Active Galactic Nucleus (AGNs), characterized by ultra relativistic jets that are oriented very close to the observer's line-of-sight, ejecting from a supermassive black hole (SMBH) that accretion makes the activity in AGNs and blazars after all, within which relativistic particles radiate losing their energy in a magnetic field \citep{UP95}. Blazars exhibiting distinctive and extreme observational properties, such as large amplitude and rapid variability, superluminal motion, high polarization, and strong emission over the entire electromagnetic spectrum \citep{Wil92, FX96, Bai98, Rom02, Fan05, Fan11, Fan16, Ghi10, Abd09, Abd10c, Abd10d, Urr11, Mar11, Yan12, Gup12, Ace15, Pei16, Xia19, 4LAC, Fan21, Bur21}. All of these properties are due to the relativistic beaming effect \citep[e.g.,][]{MGP87, Ghi93, DG95, Sav10, Fan09, Fan13, Pei19, Pei20PASP, Pei20PASA}. Blazars are the most common $\gamma$-ray-emitting objects in the extragalactic sky and also represent the most abundant population of extragalactic sources at TeV energies \citep{Hof18, Di19, 4FGL, 4LAC}.


Traditionally, based on the optical spectral features, blazars are grouped into flat-spectrum radio quasars (FSRQs) and BL Lac objects (BL Lacs) \citep{SF97}, where BL Lacs have weak or no emission lines (i.e. the equivalent width, EW, of the emission line in rest frame is less than 5 $\mathring{A}$), while FSRQs show stronger emission lines \citep[EW $\ge5\:\mathrm{\AA}$,][]{Sto91, Sti91, UP95} in their optical spectra. A more physically intuitive classification between BL Lacs and FSRQs can be distinguished based on the luminosity of the broad-line region (BLR) measured in Eddington units that the FSRQs have $L_{\rm BLR}/L_{\rm Edd}\ge5\times10^{-4}$ while BL Lacs have less than this criterion \citep{Ghi11}, implying that they have different accretion regimes \citep{Sba14}. Blazars can also be categorised via their spectral energy distributions (SEDs) synchrotron peak frequencies $\log \, \nu_{p}$ \citep{Abd10d, Gio12AA}. Low-synchrotron peaked blazars (LSPs) are characterized by $\log \, \nu_{p}$(Hz)$<$14, and intermediate-synchrotron peaked blazars (ISPs) have 14$<$$\log \, \nu_{p}$(Hz)$<$15, while $\log \, \nu_{p}$(Hz)$>$15 pertains to high-synchrotron peaked blazars (HSPs). The majority of HSPs and ISPs blazars have been classified as BL Lacs, while LSPs ones include FSRQs and some low-frequency-peaked BL Lacs \citep[see][]{Abd10d, Fan16, Bot19}. In this sense, BL Lacs can be divided into high-synchrotron peaked BL Lacs (HBLs) intermediate-synchrotron peaked BL Lacs (IBLs) and low-synchrotron peaked BL Lacs (LBLs).    

One of the goals of studying $\gamma$-ray blazars is to develop a unified framework in which two subclasses of blazars might be understood in terms of variations in a few fundamental parameters, such as the SMBH mass $M$ , the Doppler factor $\delta$, the orientation of a relativistically beamed component relative to our line-of-sight $\theta$, and the propagation angle of dispersed $\gamma$-ray emission. Manifestly, determining the masses of the central black holes of blazars is an significant step toward this goal, thus many methods have been proposed to estimate the black hole masses \citep[e.g.][]{Kas00, Xie05, Bar02, WU02, ZC09, YF10, She11, Sba12, Sha12, Pal21}. However, it should be noted that the estimation of the black hole mass for a same object from different approaches may result in a larger difference ($\sim$ two orders of magnitude in some cases).

The {\it Fermi} Gamma-ray Space Telescope with its main instrument on-board, the Large Area Telescope ( {\it Fermi}-LAT), opened a new era in the study of high-energy emission from AGNs. Many new high-energy $\gamma$-ray sources were detected, revolutionising, in particular, the knowledge of $\gamma$-ray blazars, providing us with a valuable opportunity to explore the $\gamma$-ray production mechanism. Based on the first ten years of data from the {\it Fermi} Gamma-ray Space Telescope mission, the latest catalog, 4FGL, or the fourth {\it Fermi} Large Area Telescope catalog of high-energy $\gamma$-ray sources, has been released, which includes 5778 sources above the significance of $4\sigma$, covering the 50 MeV$-$1 TeV range \citep{4FGL, 4LAC, 4FGL(2), 4LAC(2)}, about 2000 more than the previous 3FGL catalog \citep{Ace15}.  AGNs are the vast majority of sources in 4FGL; among them 3421 blazars, or 1191 BL Lacs, 733 FSRQs, and 1498 blazar candidates of unknown class (BCUs).

In this present paper, we estimate four fundamental physics parameters for $\gamma$-ray blazars which involves the upper limit of central black hole mass, the Doppler factor, the location of $\gamma$-ray region and the emission propagation angle, aiming to probe their relations and shed new light on the relativistic beaming effect and $\gamma$-ray emission mechanism of blazars detected by {\it Fermi}-LAT. The method we use was firstly proposed by \citet{BK95}, where they discussed only one applicant, 3C 279, thus we enlarge the $\gamma$-ray blazars sample and raise some forward-looking perspectives. The model we apply is going to presented in Section \ref{sec2}, while in Section \ref{sec3} we describe the sample and the derived results will be presented in Section \ref{sec4}. In Section \ref{sec5} we conduct the statistical analysis and make discussions. We draw the conclusions in Section \ref{sec6}. Throughout this paper, we adopt the $\Lambda$CDM model with $\Omega_{\Lambda}\simeq0.73$, $\Omega_{M}\simeq0.27$, and $H_{0}\simeq$ 73 km s$^{-1}$ Mpc$^{-1}$.

\section{Method} \label{sec2}

\begin{figure}
   \centering
   \includegraphics[width=12cm]{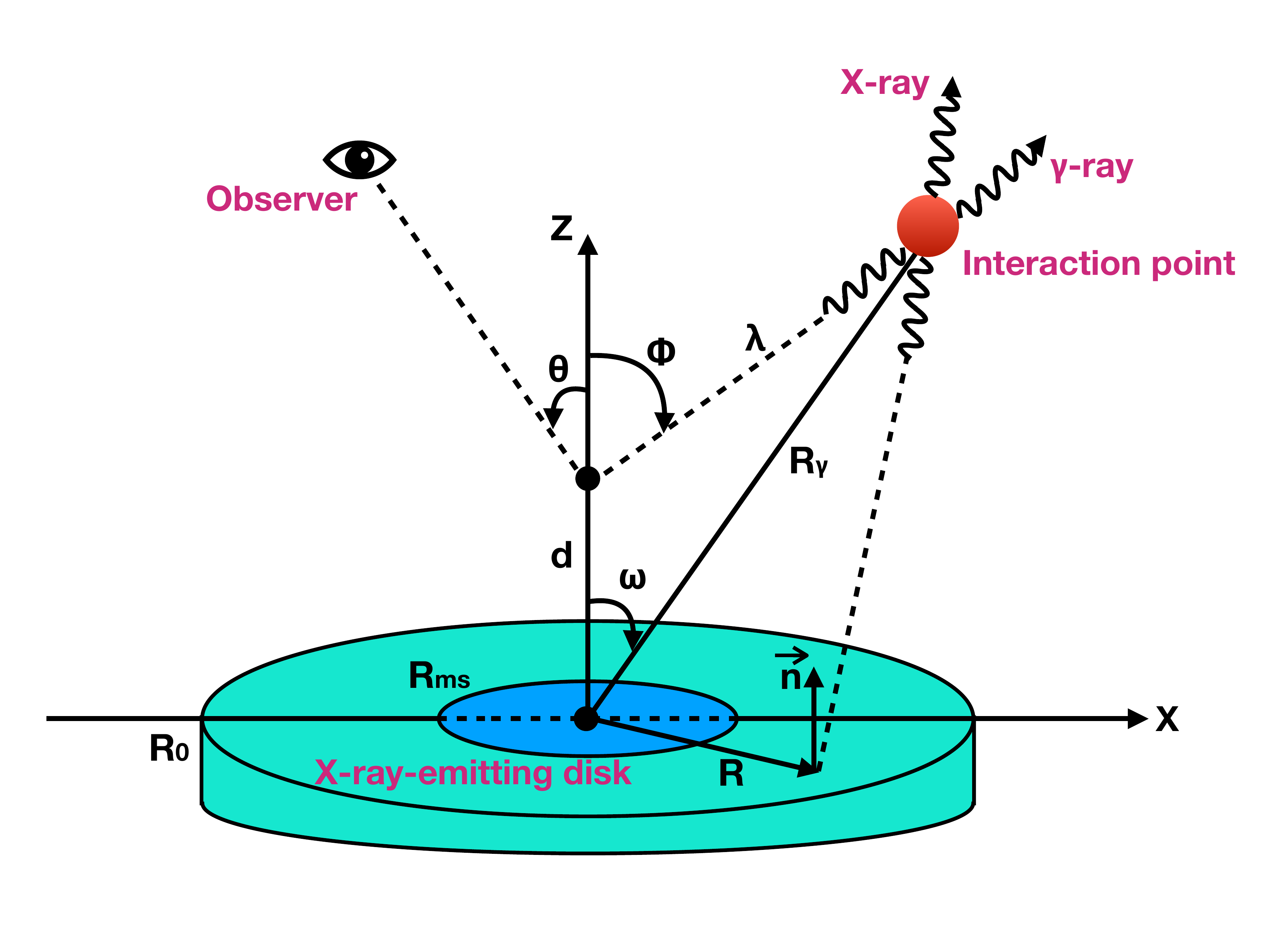}
   \caption{Illustrative schematic of the $\gamma$-rays propagation geometry above a two-temperature disk surrounding a supermassive black hole. The $\gamma$-rays interact with the soft X-ray photons produced at all points on the disk, locating at the $\gamma$-ray-emitting region of $R_{\gamma}$ from the origin with a polar angle $\omega$. $d$ represents the distance along the axis to the site of the $\gamma$-ray production and $\lambda$ is the distance traversed by the $\gamma$-ray since its creation. The propagation angle with respect to the axis of the accretion disk is $\Phi$ and $\theta$ is the viewing angle from the observer. This diagram is redrawn and modified from the original figure, see \citet{BK95} and \citet{Fan05AA}.}
     \label{fig1}
\end{figure}

It is generally believed that the escape of high energy $\gamma$-rays from AGNs depends on the $\gamma$-$\gamma$ pair production process since plenty of soft photons are surrounding the central black hole. Therefore, we can use the opacity of $\gamma$-$\gamma$ pair production  to constrain the fundamental physics parameters for $\gamma$-ray blazars. \citet{BK95}  calculated the $\gamma$-ray optical depth in the X-ray field of an accretion disk and found that the $\gamma$-rays should escape preferentially along the symmetric axis of the disk due to the strong angular dependence of the pair production cross section. The phenomenon of $\gamma$-$\gamma$ relating to the more general issue of $\gamma$-$\gamma$ transparency can set a minimum distance between the central black hole and the site of $\gamma$-ray production \citep{Bed93, DS94, Che99}. Thus the $\gamma$-rays are constrained in a solid angle, i.e., $\Omega=2\pi(1-\cos\Phi)$, yielding that the apparent observed $\gamma$-ray luminosity can be expressed as $L_{\gamma}^{\rm obs}=\Omega d_{\rm L}^{2}F_{\gamma}^{\rm obs}$, where $d_{\rm L}$ denotes the luminosity distance and $F_{\gamma}^{\rm obs}$ is the observed $\gamma$-ray flux. The observed $\gamma$-rays from AGNs require that the jet almost points to us and the optical depth $\tau$ is not greater than unity, i.e., $\tau\le1$. Since the $\gamma$-rays come from a solid angle $\Omega$ instead of being isotropic then the non-isotropic radiation, thus the absorption and beaming effects should be considered when the properties of a $\gamma$-ray-loud blazars are discussed. Besides, the variability time scale also affects the $\gamma$-ray emission region. All of all, based on these considerations, we deduce an effective method to derive four fundamental physics parameters including the upper limit of the central black hole mass ($M$), the Doppler factor ($\delta$), the distance along the axis to the site of the $\gamma$-ray production ($d$) and the propagation angle ($\Phi$) for selected {\it Fermi}-LAT detected blazars (see Figure \ref{fig1} for model elucidation).           

Since the detailed calculation process had been presented in \citet{BK95} and also in our previous papers \citep[e.g.,][]{Che99, Fan09RAA}, therefore we only list ultimate four equations here. Readers may refer to the above papers and also in Appendix \ref{App1}.   
\begin{eqnarray}
&\displaystyle\frac{d}{R_{g}}&=1730\times\frac{\Delta T_{\rm D}}{1+z}\delta M_{7}^{-1}, \nonumber \\
&L_{\rm iso}^{45}&=\displaystyle\frac{2.52\lambda\delta^{\alpha_{\gamma}+4}}{(1-\cos\Phi)(1+z)^{\alpha_{\gamma}-1}}M_{7}, \nonumber \\
&9\times&\Phi^{2.5}\left(\displaystyle\frac{d}{R_{g}}\right)^{-\frac{2\alpha_{X}+3}{2}}+kM_{7}^{-1}\left(\displaystyle\frac{d}{R_{g}}\right)^{-2\alpha_{X}-3}=1, \nonumber \\
&22.5&\times\Phi^{1.5}(1-\cos\Phi)-9\times\frac{2\alpha_{X}+3}{2\alpha_{\gamma}+8}\Phi^{2.5}\sin\Phi-\frac{2\alpha_{X}+3}{2\alpha_{\gamma}+4}kM_{7}^{-1}A^{-\frac{2\alpha_{X}+3}{2}}(1-\cos\Phi)^{-\frac{2\alpha_{X}+3}{2\alpha_{\gamma}+8}}\sin\Phi=0.
\label{eq1}
\end{eqnarray} \citep{Che99, Fan05AA, Fan09RAA}.
Here, $d$ is in units of the Schwarzschild radius $R_{g}$, $\Delta T_{\rm D}$ is the variability time scale in units of days, $z$ denotes the redshift, $L_{\rm iso}^{45}$ is the isotropy luminosity in units of $10^{45}$ erg s$^{-1}$, $\alpha_{X}$ and $\alpha_{\gamma}$ refer to the X-ray and $\gamma$-ray spectral index, respectively. $\lambda$ is a parameter depending on specific $\gamma$-ray emission model, $k$ and $A$ are coefficients (see Appendix \ref{App1}). Therefore, solving Equations (\ref{eq1}), four fundamental physics parameters, the upper limit of central black hole mass, $M_{7}$ ($=10^{7}M_{\sun}$), the Doppler factor, $\delta$, the distance along the axis to the site of the $\gamma$-ray production, $d/R_{g}$, and the propagation angle with respect to the axis of the accretion disk, $\Phi$, can be estimated from the knowledge of the redshift, $z$, and luminosity distance, $d_{\rm L}$, the X-ray behaviour (characterised by spectral index $\alpha_{X}$ and flux density), the $\gamma$-ray behaviour (characterised by spectral index $\alpha_{\gamma}$ and average $\gamma$-ray photon energy $E_{\gamma}$), and the timescale of variation $\Delta T_{D}$ are given. We adopt $\Delta T_{D}=1$ day and $\lambda=0.1$ in our calculation.

\section{Sample} \label{sec3}

Recently, \citet{Pei20PASA} compiled a sample of total 809 $\gamma$-ray blazars detected by {\it Fermi} and listed in the 4FGL catalog, with the purpose of estimating the $\gamma$-ray Doppler factor ($\delta_{\gamma}$). This approach requires the available X-ray and $\gamma$-ray emission characteristics, thus we collected 660 blazars with given X-ray data from \citet{Yan19}, which they probed the origin of X-ray emission there. For the rest of 149 sources, we sought for their X-ray data by way of NED (NASA/IPAC Extragalactic Database\footnote{http://ned.ipac.caltech.edu/}), BZCAT (The Roma BZCAT-5th edition, Multi-frequency Catalogue of Blazars\footnote{http://www.asdc.asi.it/bzcat/}) \citep{Mas15}, and other references. On the other hand, their $\gamma$-ray data are adopted from 4FGL. Therefore, we finally selected 809 {\it Fermi}-detected blazars with available X-ray and $\gamma$-ray emission characteristics to calculate the $\gamma$-ray Doppler factor. This sample contained 468 BL Lacs and 341 FSRQs. Based on the classification in \citet{Fan16}, 35 LBLs+231 IBLs+202 HBLs constitute our BL Lacs sample. In consideration of that our method given in this work also requests for the X-ray and $\gamma$-ray behaviors, thus we employ this identical sample in this work.      

Four fundamental physics parameters can be derived for these 809 sources which are comprised of 341 FSRQs and 468 BL Lacs. The overall sample and their related data are listed from Col. (1) to (8) in Table \ref{1}, which Col. (1) presents 4FGL name listed in {\it Fermi}-LAT; Col. (2) other name; Col. (3) classification; Col. (4) redshift; Col. (5) $\gamma$-ray photon index; Col. (6) $\gamma$-ray luminosity in units of erg s$^{-1}$; Col. (7) X-ray spectral index and Col. (8) flux density at 1 keV in units of $\mu$Jy.

\begin{deluxetable*}{cccccccccccc}
\tablenum{1}
\tablecaption{Sample of $\gamma$-ray blazars\label{1}}
\tablewidth{0pt}
\tablehead{
\colhead{4FGL Name} & \colhead{Other Name} & \colhead{Class} & \colhead{$z$} & \colhead{$\alpha_{\gamma}^{\rm ph}$} & \colhead{$L_{\gamma}$} & \colhead{$\alpha_{X}$} & \colhead{$F_{1\rm\, keV}$} & \colhead{$M_{7}$} & \colhead{$\delta$} & \colhead{$d/{R_{g}}$} & \colhead{$\Phi$}
}
\decimalcolnumbers
\startdata
4FGL J0004.4-4737	&	PKS 0002-478	&	FSRQ	&	0.880	&	2.42	&	46.03	&	1.42	&	0.11	&	21.66	&	1.54	&	65.24	&	10.25	\\
4FGL J0005.9+3824 	&	0003+380	&	FSRQ	&	0.229	&	2.67	&	44.46	&	1.32	&	0.08	&	17.12	&	0.81	&	66.72	&	17.15	\\
4FGL J0006.3-0620 	&	0003-066 	&	HBL	&	0.347	&	2.17	&	44.48	&	1.17	&	0.152 &	11.24	&	0.86	&	97.81	&	27.44	\\
4FGL J0008.0+4711	&	MG4 J000800+4712	&	IBL	&	0.280	&	2.06	&	45.52	&	1.05	&	0.058 &	74	&	0.9	&	11.65	&	8.95	\\
4FGL J0008.4-2339	&	RBS 0016	&	IBL	&	0.147	&	1.68	&	44.08	&	0.68	&	0.402 &	4.72	&	0.82	&	260.94	&	10.24	\\
4FGL J0010.6+2043 	&	0007+205	&	FSRQ	&	0.600	&	2.32	&	45.12	&	1.32	&	0.058 &	16.96	&	1.07	&	68.2	&	19.95	\\
4FGL J0013.9-1854	&	RBS 0030	&	IBL	&	0.095	&	1.97	&	43.66	&	0.97	&	1.026 &	7.41	&	0.63	&	133.37	&	16.71	\\
4FGL J0014.1-5022	&	RBS 0032	&	HBL	&	0.569	&	1.99	&	45.38	&	0.99	&	0.808 &	11	&	1.28	&	128.35	&	16.65	\\
4FGL J0014.2+0854 	&	0011+086	&	HBL	&	0.163	&	2.50	&	43.75	&	1.50	&	0.059 &	16.81	&	0.59	&	52.56	&	24.68	\\
4FGL J0016.2-0016	&	S3 0013-00	&	FSRQ	&	1.577	&	2.73	&	46.73	&	1.73	&	0.028 &	18.11	&	2.21	&	81.91	&	18.75	\\
$\cdots$	&	$\cdots$	&	$\cdots$	&	$\cdots$	&	$\cdots$	&	$\cdots$	&	$\cdots$	&	$\cdots$ &	$\cdots$	&	$\cdots$	&	$\cdots$	&	$\cdots$	\\
\enddata
\tablecomments{Column information is as follows: Col. (1) gives 4FGL name presented in {\it Fermi}-LAT; Col. (2) other name; Col. (3) classification (FSRQ: flat spectrum radio quasar; HBL: high synchrotron peak BL Lacs; IBL: intermediate synchrotron peak BL Lacs; LBL: low synchrotron peak BL Lacs); Col. (4) redshift; Col. (5) $\gamma$-ray photon index; Col. (6) $\gamma$-ray luminosity in units of erg s$^{-1}$; Col. (7) X-ray spectral index; Col. (8) flux density at 1 keV in units of $\mu$Jy; Col. (9) derived black hole mass in units of $10^{7}M_{\sun}$; Col. (10) derived Doppler factor; Col. (11) derived distance along the axis to the site of the $\gamma$-ray production in units of $R_{g}$ and Col. (12) derived propagation angle.\\\\
(The table is available in its entirety in machine-readable form)}
\end{deluxetable*}

\section{Results} \label{sec4}

We derive $M_{7}$, $\delta$, $d/R_{\gamma}$ and $\Phi$ for every single sources in our sample using Equations (\ref{eq1}), and the results are presented in the last four columns in Table \ref{1}.

\begin{figure}
   \centering
   \includegraphics[width=8cm]{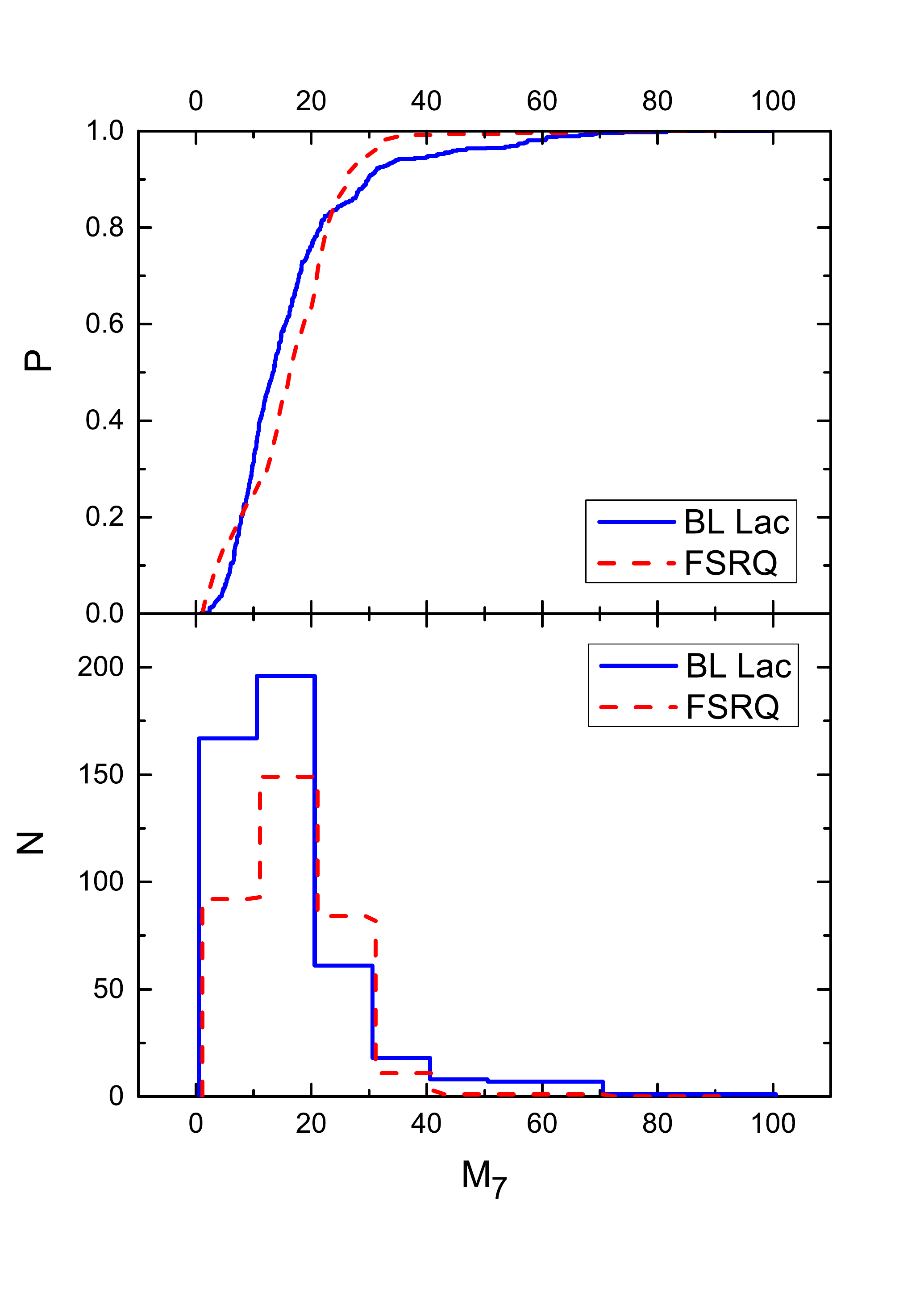}
   \includegraphics[width=8cm]{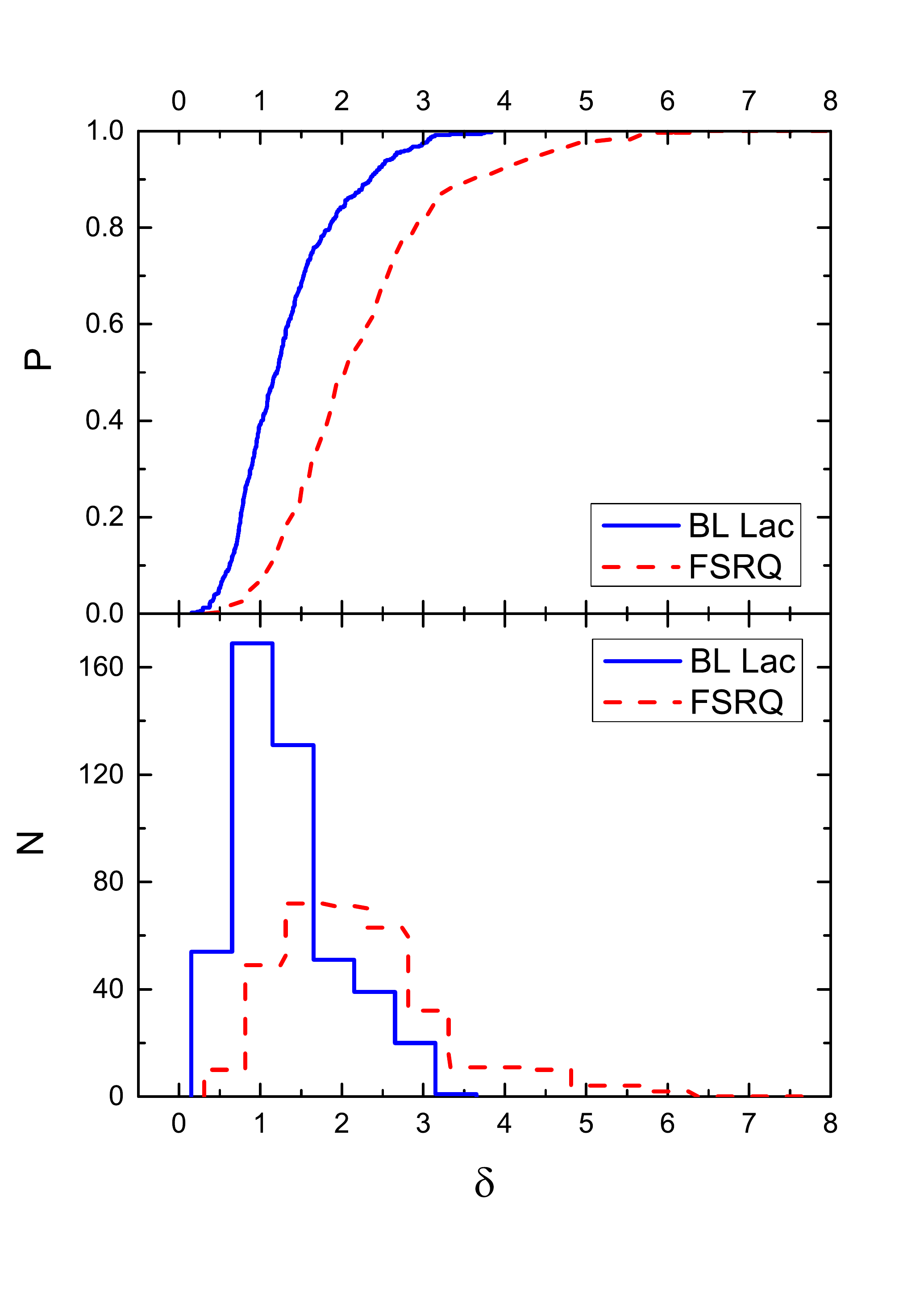}
   \includegraphics[width=8cm]{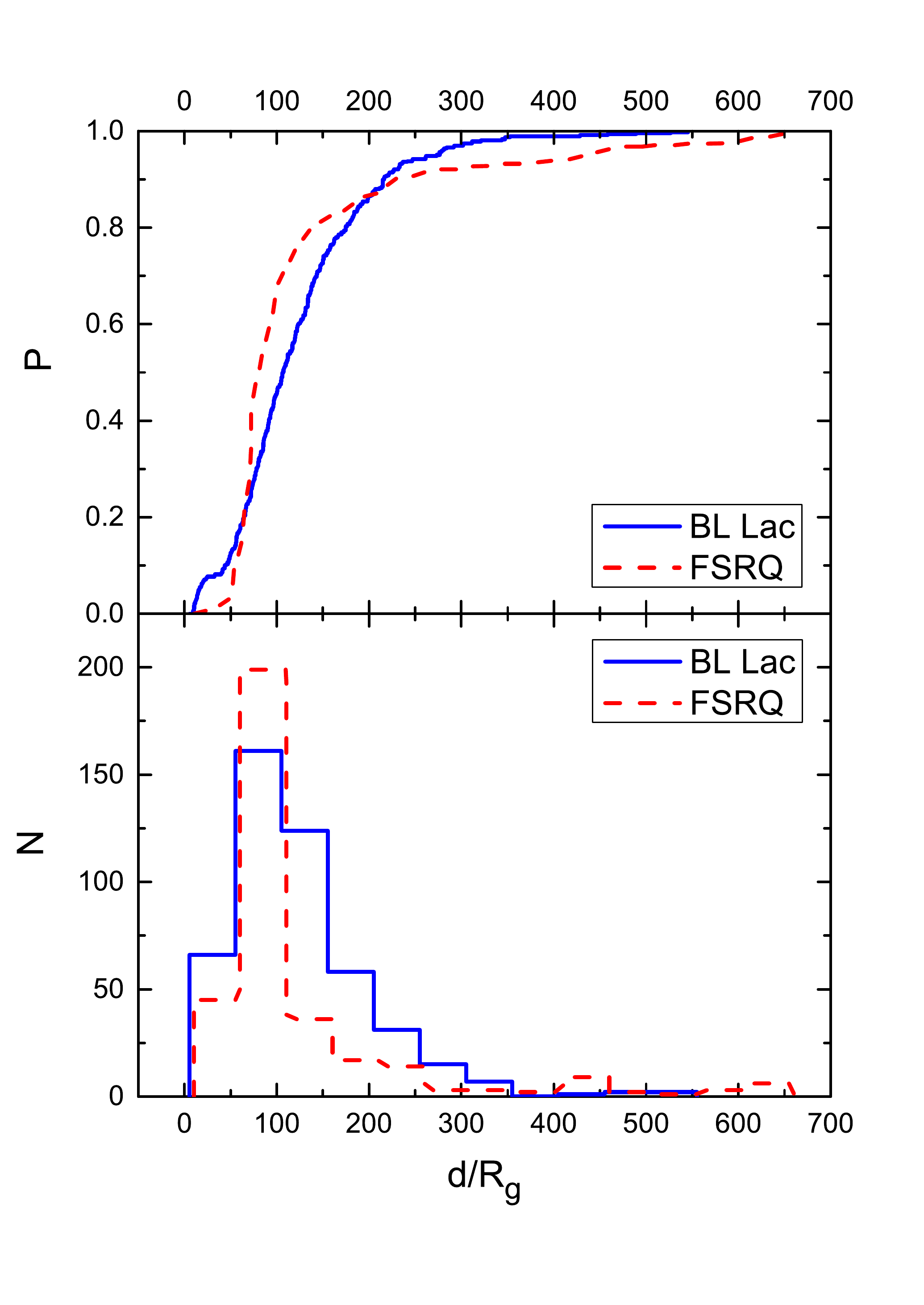}
   \includegraphics[width=8cm]{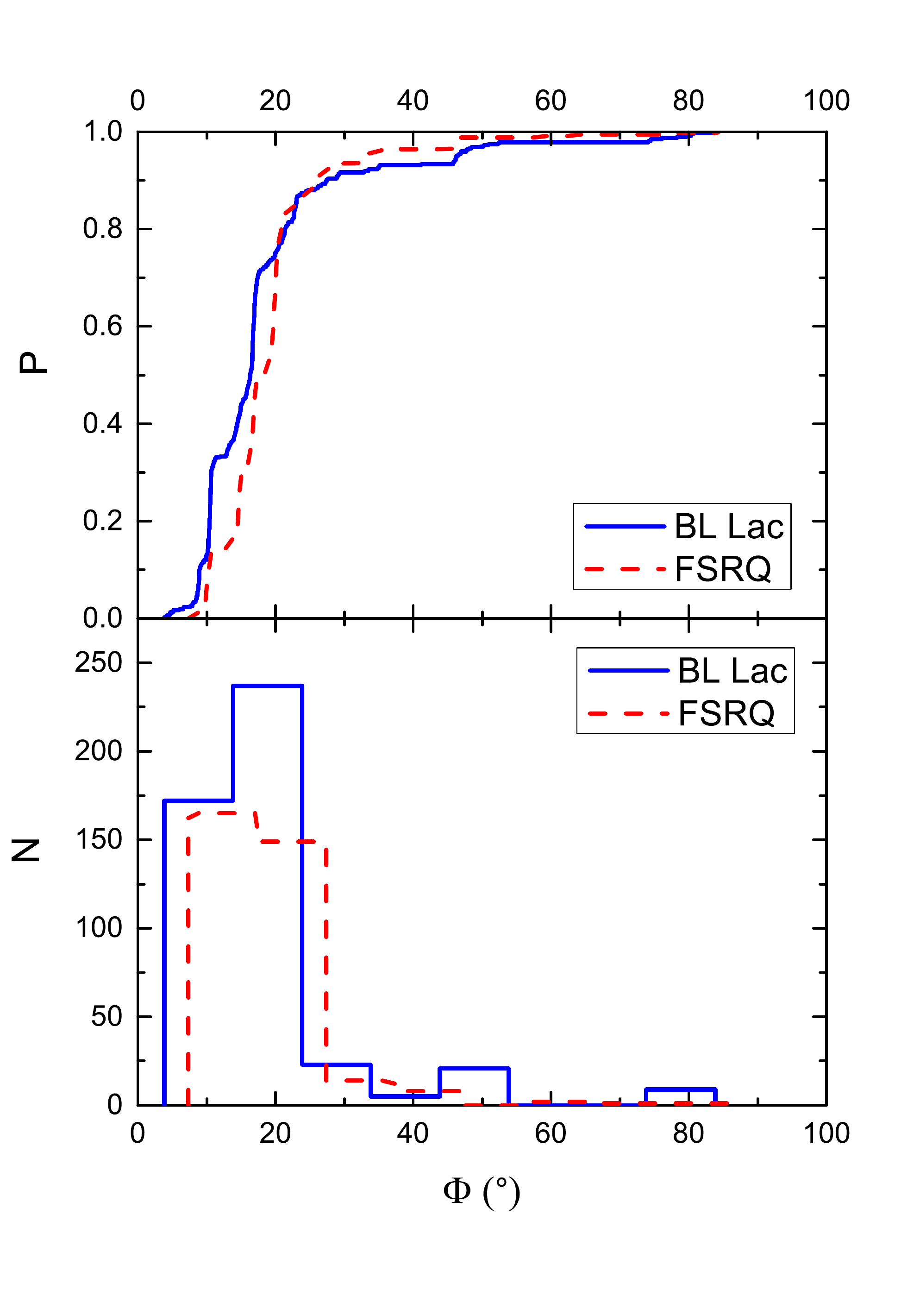}
   \caption{Distributions of the upper limit of central black hole mass ($M_{7}$), Doppler factor ($\delta$), the distance along the axis to the site of the $\gamma$-ray production ($d/R_{g}$), and the propagation angle with respect to the axis of the accretion disk ($\Phi$) for BL Lacs and FSRQs. In this Figure, the {\it red solid line} stands for BL Lacs and {\it blue dashed line} for FSRQs.}
     \label{fig2}
\end{figure}

The upper left panel in Figure \ref{fig2} shows the distributions of $M_{7}$ for BL Lacs and FSRQs. The ranges are from 0.54 to 99.90 with an average value of $16.56\pm12.74$ and a median of 13.34 for 468 BL Lacs, and from 1.11 to 91.46 with an average value of $16.38\pm9.77$ and a median of 16.24 for 341 FSRQs. A Kolmogorov-Smirnov test (hereafter K-S test) is performed on two sub-samples and we obtain that the null hypothesis (they both are from the same population) cannot be rejected at the confidence level $P=4.48\times10^{-5}$ ($d_{\rm max}=0.16$) for BL Lacs and FSRQs. Thus at the 0.0001 level, these two distributions are different. However, we can find a rough overlap between 0 to 30 from the histogram. If we slightly fine-tune the confidence level at $10^{-5}$, then two classes likely belong to the same parent distribution, suggesting that the central black hole mass perhaps plays a less important role in the evolutionary sequence of blazars \citep[e.g.][]{WLZ02, BD02}.      

With respect to BL Lacs, we obtain medians of 10.86, 15.31 and 13.72 for HBLs, IBLs and LBLs separately, implying that HBLs may hold a lighter black hole mass.      

The distributions of $\delta$ for BL Lacs and FSRQs are displayed in the upper right panel in Figure \ref{fig2}, spanning from 0.15 to 3.84 with an average value of $1.32\pm0.67$ and a median of 1.20 for BL Lacs, and from 0.31 to 7.96 with an average value of $2.24\pm1.10$ and a median of 2.03 for FSRQs. The K-S test is carried out and we ascertain $P=4.36\times10^{-36}$ and $d_{\rm max}=0.45$. This extremely small $P$-value indicates that the two classes significantly come from a different parent population and thus Doppler factor for FSRQs is on average higher than that for BL Lacs. On the other hand, severally, medians of 1.15, 1.22 and 1.26 for 202 HBLs, 231 IBLs and 35 LBLs are also acquired. This appearing sequence that HBL $\sim$ IBL $\sim$ LBL $\sim$ FSRQ of $\delta$ supports previous findings \citep[e.g.,][]{Mar08MN, Ghi10, Ghi17MN, Xio15a, Xio15b, Ghi16, RC16} and also our previous conclusions \citep{Fan13RAA, Pei20PASA, Pei20RAA}, revealing that Doppler effect varies in different sub-classes of blazars.      
 
The lower left panel in Figure \ref{fig2} presents the distributions of $d/R_{g}$ for our sample, which are in the scope from 5.12 to 545.66 with a median of 106.48 for BL Lacs, and from 10.21 to 655.97 with a median of 80.02 for FSRQs. The K-S test reports that $P=9.00\times10^{-9}$ and $d_{\rm max}=0.22$.  

Finally, the distributions of $\Phi$ for BL Lacs and FSRQs are exhibited in the lower right panel in Figure \ref{fig2}. The values are in the extent between 3.84$^\circ$ and 83.97$^\circ$ with a median of 16.33$^\circ$ for BL Lacs, and from 7.33$^\circ$ to 84.31$^\circ$ with a median of 18.08$^\circ$ for FSRQs. $P$-value of $2.26\times10^{-9}$ ($d_{\rm max}=0.23$) is shown from K-S test. 

We summary our derived results and distribution statistics on four parameters in Table \ref{2}. Considering all the K-S test results, we can obtain the fact that BL Lacs and FSRQs belong to different parent distributions of evolution.

\begin{deluxetable*}{cccccccc}
\tablenum{2}
\tablecaption{Descriptive statistics of the derived fundamental physics parameters for the whole sample\label{2}}
\tablewidth{5pt}
\tablehead{
    & \multicolumn3c{BL Lacs} & \multicolumn1c{} & \multicolumn3c{FSRQs}}
\startdata
N & \multicolumn3c{468} & \multicolumn1c{} & \multicolumn3c{341} \\
Parameter & Median & Maximum & Minimum & K-S test & Median & Maximum & Minimum \\ \hline
$M_{7}$ & 13.34 & 0.54 & 99.90 & $P=4.48\times10^{-5}$ ($d_{\rm max}=0.16$) & 16.24 & 1.11 & 91.46 \\
$\delta$ & 1.20 & 0.15 & 3.84 & $P=4.36\times10^{-36}$ ($d_{\rm max}=0.45$) & 2.03 & 0.31 & 7.96 \\
$d/R_{g}$ & 106.48 & 5.12 & 545.66 & $P=9.00\times10^{-9}$ ($d_{\rm max}=0.22$) & 80.02 & 10.21 & 655.97 \\
$\Phi$ ($^{\circ}$) & 16.33 & 3.84 & 83.97 & $P=2.26\times10^{-9}$ ($d_{\rm max}=0.23$) & 18.08 & 7.33 & 84.31 \\
\enddata
\end{deluxetable*}

\section{Discussion} \label{sec5}

\subsection{Black hole mass, $M_{7}$} \label{sec5.1}

The central black hole plays an important role in the observational properties of AGNs and has drawn much attention. It may also shed some new light on the evolution process \citep[e.g.,][]{Bar02}. There are several methods for the estimations of black hole mass. Traditionally, the virial black hole mass can be estimated by adopting an empirical relationship between broad line region (BLR) size and ionizing luminosity combined with the measured broad-line width, which is assumed that the BLR clouds are gravitationally bound by the central black hole with Keplerian velocities. This traditional virial method for estimating the black hole mass is usually applied in FSRQs \citep{Shen11, Sba12, Sha12}. The black hole mass for BL Lacs can be estimated from the properties of their host galaxies namely $M-\sigma$ or $M-L_{\rm bul}$ relations since BL Lacs have no or weak emission line. Here $\sigma$ and $L_{\rm bul}$ refer to the stellar velocity dispersion and the bulge luminosity of the host galaxies \citep{Bar02, WU02, ZC09, Chai12}. Other idiomatic estimated methods such as the reverberation mapping \citep[e.g.][]{Kas00} and variability time-scale approaches \citep[e.g.][]{Fan99, YF10} are usually applied for the black hole mass determinations although consensus has not been reached.

In this present paper, we enlarge the sample of {\it Fermi}-detected blazars with derived black hole masses $M_{7}$ following the idea from previous studies \citep{Che99, Fan05, Fan09RAA}. This estimated method is constrained by the optical depth of the $\gamma-\gamma$ pair production. We need to point out that the determined black hole mass is an upper limit due to the restriction on the optical depth of unity. It also should be noted that the main difference between our calculation and others lies on we consider the $\gamma$-rays originate from a cone with a solid angle $\Omega=2\pi(1-\cos\Phi)$ and others assume that $\gamma$-rays are isotropic, i.e., $\Omega=4\pi$.            

\begin{figure}
   \centering
   \includegraphics[width=12cm]{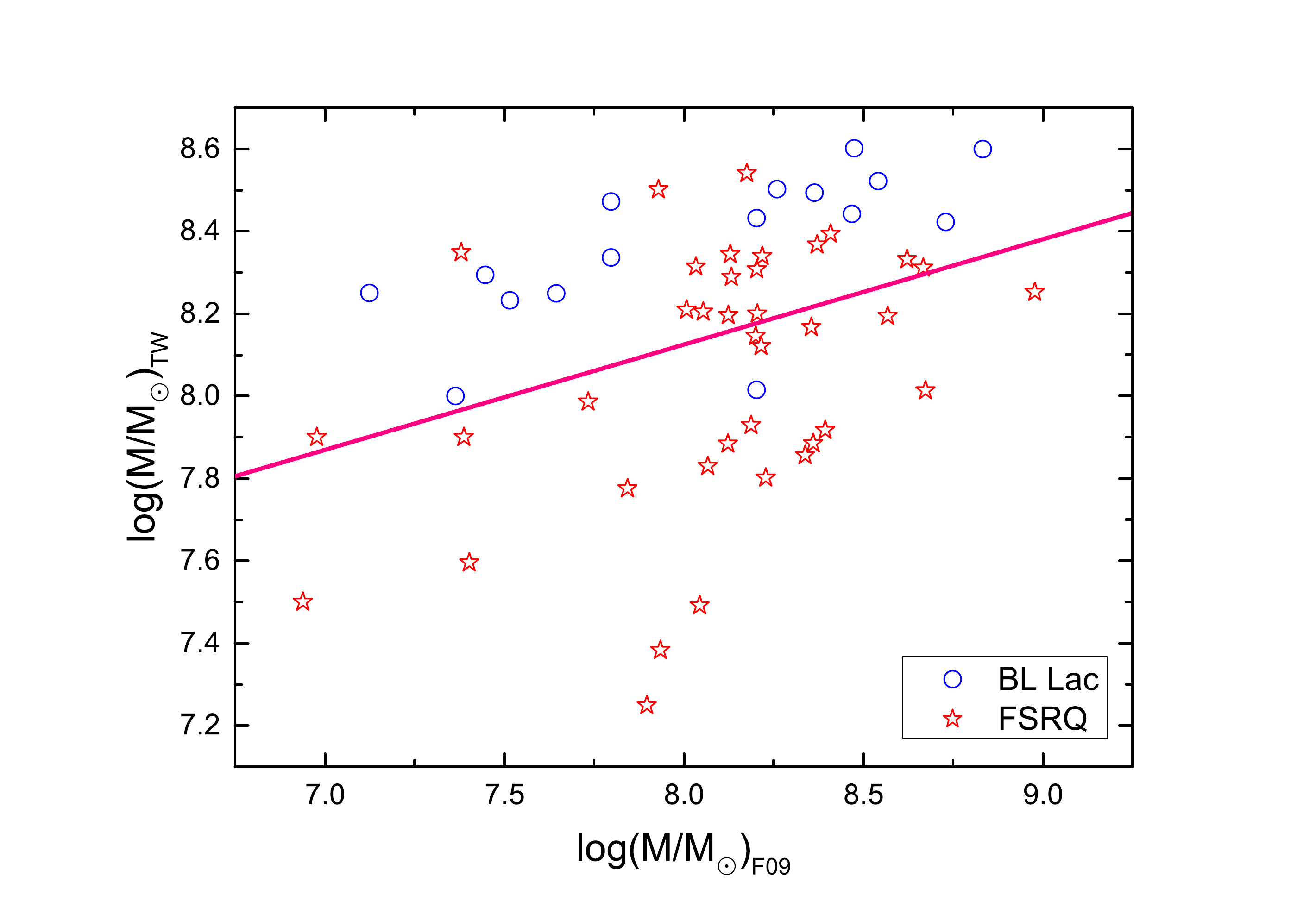}
   \caption{Plot of the correlation between the estimated black hole mass $\log(M/M_{\odot})$ derived from this paper (denoted by a subscript `TW') and from \citet{Fan09RAA} (denoted by a subscript `F09').}
     \label{fig6}
\end{figure}

For verifying the conformance of our results with the previous work, we cross-checked our sample with \citet{Fan09RAA}, and found all there 54 sources are included in our sample\footnote{In fact, \citet{Fan09RAA} studied 59 $\gamma$-ray blazars, but 5 of them are not listed in 4FGL.}. They found the average values of $M_{7}$ are 13.18 for BL Lacs and 11.75 for FSRQs in their sample. In this present paper, we obtain the average values are 16.56/16.38 for 468 BL Lacs/ 341 FSRQs respectively, which is consistent with \citet{Fan09RAA} showing that the difference of black hole mass in BL Lacs and FSRQs are not large. Figure \ref{fig6} displays the plot of those 54 cross-checked sources and there is well correlated between two groups data with correlation coefficient $r=0.37$ and a chance probability of $P=0.006$.    

In despite of showing coherence in estimation of $M_{7}$ between two samples, we intend to remark some calculative differences within the same method. This method has been firstly proposed in \citet{Che99}, where they investigated 7 $\gamma$-ray loud blazars. Afterwards, \citet{Fan05} and \citet{Fan09RAA} proceed this calculation to a larger selected sample of $\gamma$-ray blazars. Using the estimation kernel and in the light of Equation (\ref{eq1}), we can derive $M_{7}$ and other three fundamental parameters if the knowledge of the cosmological behaviour characterising by redshift $z$ and luminosity distance $d_{\rm L}$, X-ray behaviour characterising by X-ray spectral index $\alpha_{X}$ and flux density at 1 keV, $\gamma$-ray behaviour characterising by $\gamma$-ray spectral index $\alpha_{\gamma}$, $\gamma$-ray flux and averaged $\gamma$-ray photon energy $E_{\gamma}$, and the timescale of variation $\Delta T$ are given. In the previous work, they all provided $\Delta T$ for each source. For instance, \citet{Che99} gave variability timescales ranging from 3.2 to 24 hours for each of 7 sources, \citet{Fan05} laid out $\Delta T$ for 23 blazars, from 1.92 to 144 hours. However, for our large sample, the variability timescales for most sources are unknown or given several values by different literatures. Since many authors had pointed out that a typical timescale of variation in the source frame for {\it Fermi}-detected blazars is around 1 day \citep{LAT11, Bon11, Nal13, Hu14, Zha15, Fan16, Che18, Pri20}. Therefore, for the sake of simplicity, we apply $\Delta T= 1$ day for all sources in our present calculation.             

Secondly, previous studies have adopted averaged photon energy $E_{\gamma}=1$ GeV uniformly. Here we calculate $E_{\gamma}$ by $E_{\gamma}=\int EdN/\int dN$, and obtain different values for each source, which are in the range from 2.01 to 9.42 GeV. To sum up, although we take different considerations, the discrepancies are negligible.   



\begin{figure}
   \centering
   \includegraphics[width=12cm]{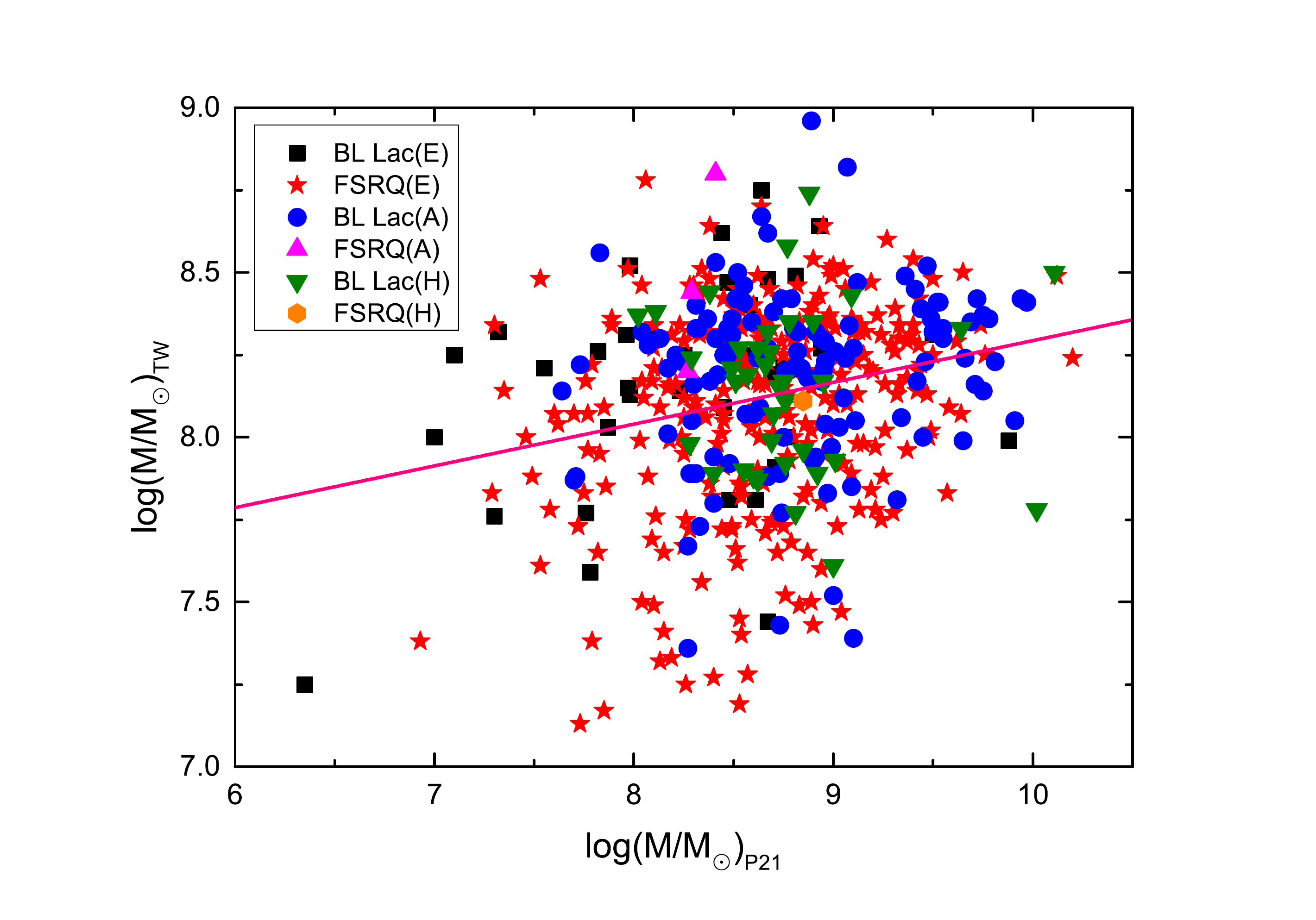}
   \caption{Plot of the correlation between the estimated black hole mass $\log(M/M_{\odot})$ derived from this paper (denoted by a subscript `TW') and from \citet{Pal21} (denoted by a subscript `P21'). The cross-checked sub-sample ascertained from BLR property, stellar velocity dispersion and Host Galaxy are labelled by `(E)', `(A)' and `(H)' after `BL Lacs' or `FSRQs', respectively. For the sake of clarity, we only present the best-fitting for the whole cross-checked sample in {\it pink solid} line, which corresponds to $\log(M/M_{\odot})_{\rm TW}=(0.13\pm0.02)\log(M/M_{\odot})_{\rm P21}+(7.03\pm0.21)$ with correlation coefficient $r=0.233$ and a chance probability of $P=3.27\times10^{-7}$.}
     \label{fig8}
\end{figure}

\citet{Pal21} presented a catalog of the central engine properties of 1077 selected {\it Fermi} blazars. They obtain the average black hole mass $M$ for the whole sample population is $\left<\log(M/M_{\odot})\right>=8.60$, which is close to our estimation with a median value of $\log(M/M_{\odot})=8.16$ in this work. Primarily, \citet{Pal21} applied three methods to compute the black hole mass. In particular, 684 sources are used BLR properties of their emission-line, 346 are adopted from stellar velocity dispersion, i.e., absorption-line, and 47 are derived from their host galaxy bulge luminosity. We cross-check our sample with their, 189 BL Lacs and 279 FSRQs are found in common. We plot the population presenting in Figure \ref{fig8}. The cross-checked sub-sample derived from the BLR property, stellar velocity dispersion and host galaxy are labelled by `(E)', `(A)' and `(H)' after `BL Lacs' or `FSRQs', respectively.        

The best-fitting for our whole cross-checked sample is $\log(M/M_{\odot})_{\rm TW}=(0.13\pm0.02)\log(M/M_{\odot})_{\rm P21}+(7.03\pm0.21)$, showing that our result is consistent with the estimation from different methods for a large sample of {\it Fermi}-detected blazars. We present this best-fitting in Figure \ref{fig8} labelled by a {\it pink solid} line. Besides, we also examine the correlations for each sub-sample derived from different estimated methods. We found a significant correlation regarding the cross-checked sub-sample using BLR luminosity (denoted by `(E)' in Figure \ref{fig8}) and the best-fitting indicates $\log(M/M_{\odot})_{\rm TW}=(0.14\pm0.03)\log(M/M_{\odot})_{\rm P21}+(6.87\pm0.26)$ with $r=0.26$ and $P=4.47\times10^{-6}$. For the sake of clarity, we do not draw this line in the plot. Whereas, we obtained unsatisfactory regressions for sub-sample adopting from stellar velocity dispersion and host galaxy, $P=0.146$ and $P=0.826$ are reported. This analysis suggests that our estimation is in a better agreement with the result derived from BLR property compared to the other two diagnoses.                  

The jet power for blazars is believed to be the order of $\dot{M}_{\rm in}c^{2}$. For BL Lacs, $\dot{M}_{\rm in}$ can be calculated by $\dot{M}_{\rm in}=P_{\rm jet}/c^2$, where $P_{\rm jet}$ is the jet power. Whereas for FSRQs, $\dot{M}$ is given by $\dot{M}=L_{\rm Disk}/\eta c^2$ with the accretion disk luminosity $L_{\rm Disk}$ and $\eta=0.08$ \citep[see detailed discussion in][]{GT08}. Then one can obtain the ratio $\dot{M}_{\rm in}/\dot{M}_{\rm Edd}$ written by 
\begin{equation}
\displaystyle\frac{\dot{M}_{\rm in}}{\dot{M}_{\rm Edd}}=\frac{\dot{M}_{\rm in}c^{2}}{1.3\times10^{38}(M/M_{\odot})}.
\label{eq20}
\end{equation}

By means of the first 3-month survey of {\it Fermi} \citep[i.e., 1FGL,][]{Abd09}, \citet{Ghi10} studied 85 sources, modeled their SEDs, and obtained black hole mass, location of the dissipation region, bulk Lorentz factor, jet power and other important physics parameters regarding those sources. They explored the distribution of $\dot{M}_{\rm in}/\dot{M}_{\rm Edd}$ and found a clear division between BL Lacs and FSRQs which took place in $\dot{M}_{\rm in}/\dot{M}_{\rm Edd}\sim0.1$. This boundary can also be expressed by $L_{\rm Disk}/L_{\rm Edd}\sim0.001$ since $L_{\rm Edd}=1.3\times10^{38}(M/M_{\odot})$ erg s$^{-1}$. This proposal of new division between two sub-classed of blazars was re-examined by \citet{CG19}. They compiled a sample including 24 BL Lacs and 77 FSRQs with available $L_{\rm Disk}/L_{\rm Edd}$. The dividing line locating in $L_{\rm Disk}/L_{\rm Edd}\sim0.01$ was also significantly discovered.

If we consider the BLR luminosity is approximately $10\%$ of the disk luminosity, i.e., $L_{\rm BLR}\simeq0.1L_{\rm Disk}$ \citep{Smi81, Cal13, CG19}, this division has further evolved to $L_{\rm BLR}/L_{\rm Edd}=5\times10^{-4}$ set by \citet{Ghi11} according to the relation between BLR luminosity and disk luminosity both measured in Eddington units. They proposed that the division of blazars occurs for a change in the accretion regime. However, the number of sources in their sample is relatively small for a strong claim (only containing 32 blazars).                         

\begin{figure}
   \centering
   \includegraphics[width=15cm]{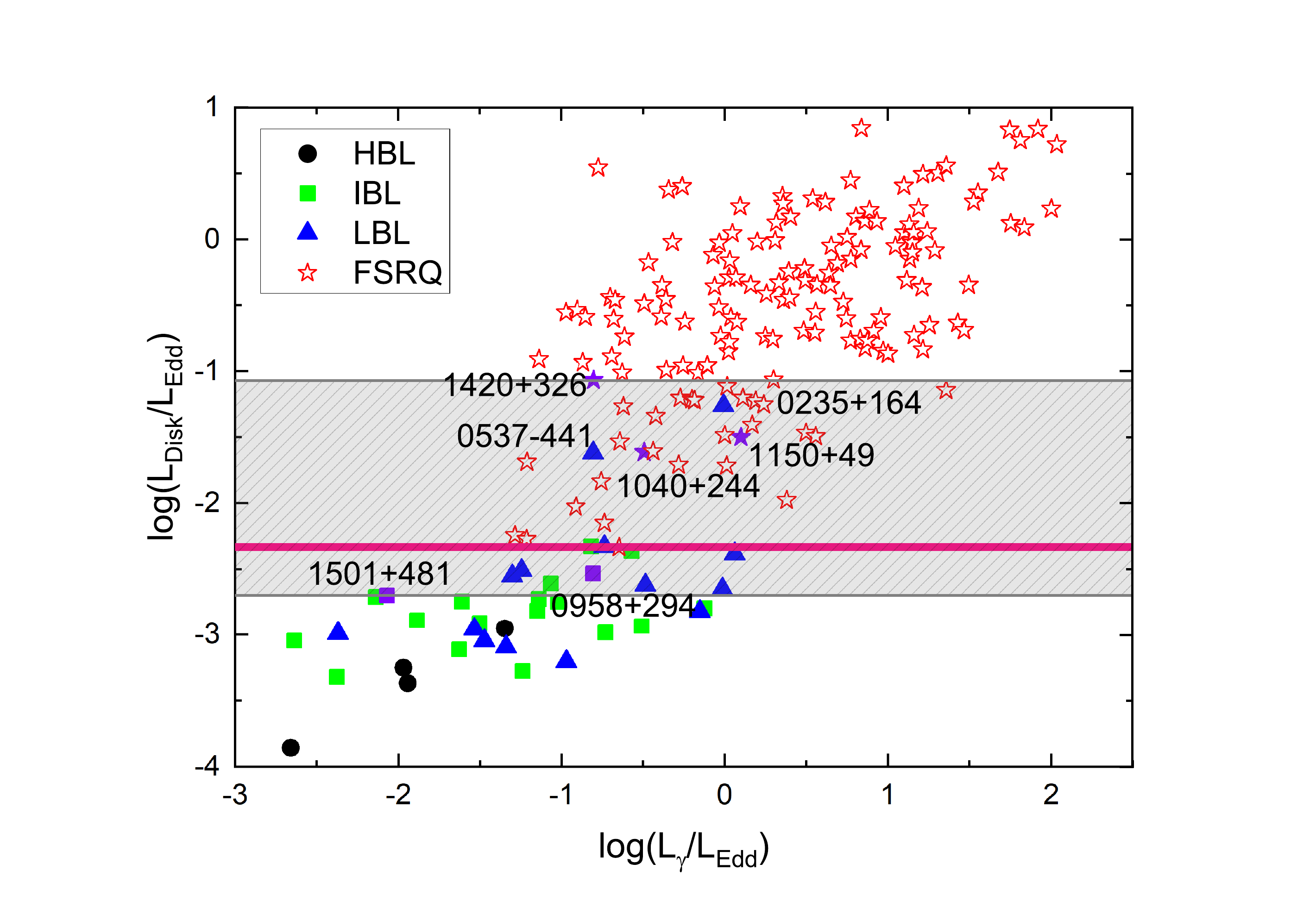}
   \caption{Plot of the BLR luminosity in units of the Eddington one versus the $\gamma$-ray luminosity in units of the Eddington one. Black circles, green squares, blue triangles and red stars denote the HBLs, IBLs, LBLs and FSRQs, respectively. The pink line locating in $\log(L_{\rm Disk}/L_{\rm Edd})=-2.33$ (i.e., $L_{\rm BLR}/L_{\rm Edd}=4.68\times10^{-4}$) is the demarcation we set in this paper to divide BL Lacs and FSRQs. Two outliers, 0235+164 and 0537-441, are perhaps masquerading BL Lacs, that is, intrinsically FSRQs with luminous accretion disk and dissipation regions within the hidden broad lines. The grey shadowed area lying on $\log(L_{\rm Disk}/L_{\rm Edd})=-2.70\sim-1.07$ (i.e., $L_{\rm BLR}/L_{\rm Edd}\simeq2.00\times10^{-4}\sim8.51\times10^{-3}$) denotes a so-called `appareling zone' signifying the potential transition field between BL Lacs and FSRQs. Five confirmed changing-look blazars locate in this zone and are labelled with {\it violet fill colour}.}
     \label{fig5}
\end{figure}

In this work, we have derived black hole masses for 809 {\it Fermi} blazars, then we can calculate their Eddington luminosities via $L_{\rm Edd}=1.3\times10^{38}(M/M_{\odot})$. Because we now have a larger sample, we can more accurately determine the dividing line on the ratio $L_{\rm Disk}/L_{\rm Edd}$ between BL Lacs and FSRQs, and also confirm the idea that the blazars' divide occur for the alteration in accretion regime \citep{Ghi09, Ghi11}. To do so, first step we search the references and collect the available BLR luminosity regarding our sample as many as possible, and finally we compile 184 sources with $L_{\rm BLR}$ (4 HBLs$+$19 IBLs$+$14 LBLs$+$147 FSRQs), which 164 sources from paper \citet{Zha20} and 20 from \citet{CG19}. Secondly, we calculate the disk luminosity $L_{\rm Disk}$. Again, we adopt $L_{\rm Disk}\simeq10 L_{\rm BLR}$. We make the plot of the BLR luminosity as a function of the $\gamma$-ray luminosity both in units of the Eddington luminosity in Figure \ref{fig5}.       

The first thing to catch our sight is a distinct tendency showing that blazars with stronger emission lines are more luminous as well in the $\gamma$-ray band. This plot also indicates that the distribution of HBLs $\to$ IBLs $\to$ LBLs $\to$ FSRQs could be explicated as a sequence of strength the broad lines. This evolution reveals the fact that some FSRQs are of the LBLs and in a few cases, IBLs. Secondly and the most importantly, the FSRQs and BL Lacs are distinguished significantly in this plot. Utilizing a Bayesian Information Criterion (BIC) and maximum likelihood estimation via the expectation-maximization (EM) algorithm \citep{FR02, Scr16}\footnote{We perform the operation by using a public domain R statistical package namely {\it mclust} available at \url{https://mclust-org.github.io/mclust/}, which provides iterative EM methods for maximum likelihood clustering with parameterized Gaussian mixture models.}, we have obtained a dividing line between these two subclasses occurring at $L_{\rm Disk}/L_{\rm Edd}=4.68\times10^{-3}$, equivalent to $L_{\rm BLR}/L_{\rm Edd}=4.68\times10^{-4}$ or $\dot{M}_{\rm in}/\dot{M}_{\rm Edd}\simeq0.0468$. This demarcation is in good agreement with the division proposed by \citet{Ghi10} or \citet{Ghi11}, and elucidating that BL Lacs turn into radiatively inefficient when $L_{\rm Disk}/L_{\rm Edd}<4.68\times10^{-3}$ whereas FSRQs show radiatively efficient when $L_{\rm Disk}/L_{\rm Edd}>4.68\times10^{-3}$ which indicates that FSRQs may have strong accretion disk.          

Remarkably, there are two outliers of BL Lacs, PKS 0235+164 (=4FGL J0238.6+1637) and PKS 0537-441 (=4FGL J0538.8-4405). These two sources are labeled as LBLs but we report the values of $L_{\rm Disk}/L_{\rm Edd}=0.055$ and $0.024$ for these two LBLs, respectively, thus they are well into the FRSQs region. We consider these two sources are potential changing-look blazars, which the variations are so dramatic that they lead to a change in classification \citep{Mat03, Sha12, Cut14, Lam15, YW18, Mis21, Pen21}. Changing-look blazars are crucial to upend our understanding of the SMBH accretion state transition and the particle acceleration process within the radio jet, which can provide us with valuable insight into AGNs and galaxies evolution. 

Therefore, these two sources are perhaps masquerading BL Lacs, i.e., intrinsically FSRQs with luminous accretion disk and dissipation regions within the hidden broad lines. A similar scenario has been proposed by \citet{Pad19} to illustrate TXS 0506+056, the first cosmic neutrino source, is not a BL Lac but is instead an FSRQ. They provided one of the evidence for re-classification that is based on its Eddington ratio. They presented $L_{\rm disk}/L_{\rm Edd}\sim0.01$ for TXS 0506+056 and therefore should be further classified as an FSRQ according to the original criterion from \citet{Ghi11}. Notably, in the same paper, \citet{Ghi11} also proposed these two sources, 0235+164 and 0537-441, are `intruder' BL Lacs and re-classified as FSRQs since SEDs modellings performed turn out that their high energy peak dominates the bolometric output and the X-ray spectrum belongs to the high-energy peak.          

\citet{Pen21} carry out an extensive search for optical spectra available in the Large Sky Area Multi-object Fibre Spectroscopic Telescope (LAMOST) Data Release 5 (DR5) archive, and discover 26 changing-look blazars. After cross-checked, there are 3 sources are also listed in our sample. They are B0958+294 (=4FGL J10001.1+2911), TXS 1501+481 (=4FGL J1503.5+4759) and TXS 1040+244 (=4FGL J1043.2+2408). The first two sources are classified as IBLs and the last one is an FSRQs in 4FGL. We label them in colour {\it violet} (see Figure \ref{fig5}). According to LAMOST investigation, \citet{Pen21} reclassified B0958+294 to be an FSRQs as well as TXS 1501+481, and TXS 1040+244 are reported as a BL Lac. 

\citet{Mis21} present multi-wavelength photometric and spectroscopic monitoring observations of the blazar, TXS 1420+326 (=4FGL J1422.3+3223), focusing on its outbursts in 2018-2020. This source is also included in our sample and originally classified as a FSRQ by {\it Fermi}-LAT. However, \citet{Mis21} found this source transitioned between BL Lac and FSRQ states multiple times following a series of flares. We also label it in colour {\it violet}.    

Based on abundant data and results from simultaneous and coordinated $\gamma$-ray and multiwavelength observations, \citet{Cut14} studied a core-dominated and radio-loud FSRQ, 4C+29.22, also known as S4 1150+49 or 4FGL J1153.4+4931, located at $z=0.334$ \citep{Ste01}. The $\gamma$-ray data in their paper were collected in the first 3 years of {\it Fermi} science observations. They found that this source showing a shift of two orders of magnitude in the frequency of the synchrotron peak (from $\sim10^{12}$ to $\sim10^{14}$ Hz) during the GeV $\gamma$-ray flare, and also displaying that an unusual flat X-ray SED that the marked spectral softening of the X-ray spectrum. All of these imply that 4C+29.22 is a typically BL Lacs, suggesting a probable transition occurs in the division of blazars. This changing-look source is also listed in our present sample.     

Overall, we found 5 confirmed changing-look blazars which are contained in our sample. The lowest value of $\log(L_{\rm Disk}/L_{\rm Edd})$ for these 5 sources is $-2.70$ (i.e., $L_{\rm BLR}/L_{\rm Edd}\simeq2.00\times10^{-4}$) given by 1501+481, and the largest values is $-1.07$ (i.e., $L_{\rm BLR}/L_{\rm Edd}\simeq8.51\times10^{-3}$) reported from 1420+326. We propose that this area falling in $\log(L_{\rm Disk}/L_{\rm Edd})=-2.70\sim-1.07$ is a so-called `appareling zone', the potential transition field between BL Lacs and FSRQs (see the grey shadowed area in Figure \ref{fig5}). The sources descending into this `appareling zone' are perhaps changing-look blazars and the transition of BL Lacs $\rightleftharpoons$ FSRQs would occur. The two outliers mentioned above, 0537-441 and 0235+164, are locating in this zone as well.

Many authors have explored the possible scenarios for this peculiar transitional phenomenon. Apart from the explanation that luminous accretion disk and dissipation regions conceal the broad lines for some FSRQs, \citet{Ghi12} also proposed that these FSRQs are with weak radiative cooling so that their broad lines are overwhelmed by nonthermal continuum. Besides, the highly beamed jets as well as the variations of jets bulk Lorentz factor, and the radiatively efficient accretion are also account for changing-look in the broad-line sources \citep{Bia09, Gio12, Rua14}.                    

To sum up, our result on a new demarcation between FSRQs and BL Lacs in terms of the accreting mass rate, that is, $L_{\rm disk}/L_{\rm Edd}=4.68\times10^{-3}$, based on a larger sample including 184 {\it Fermi} blazars, are in good agreement with the idea that the presence of strong emitting lines is in matter of conversion in the accretion regime. We also put forward that those two outliers are possibly FSRQs but showing as BL Lacs objects in disguise. Finally, we propose a `appareling zone' that BL Lacs can transit into FSRQs and vice versa, which predict that the objects locating in this zone are potentially changing-look blazars.

\subsection{Doppler factor, $\delta$}

Traditionally, the Doppler factor can be expressed by $\delta=\left[\Gamma_{\rm var}\left(1-\beta\cos\theta\right)\right]^{-1}$, where $\Gamma_{\rm var}$ is the macroscopic bulk Lorentz factor defined by $\Gamma_{\rm var}=1/\sqrt{1-\beta^{2}}$, $\beta$ is the jet speed in units of the speed of light and $\theta$ is the viewing angle between the jet and the line-of-sight. The Doppler factor $\delta$ is a crucial parameter in the jet of blazars and leading us to probe the beaming effect. However, it is difficult for us to determine this parameter since it is undetectable. Hence, some feasible methods need to be proposed \citep{Ghi93, LV99, Fan09, Hov09, Lio18, Zha20, Pei20PASA}.  


In this present work, we ascertain the derived Doppler factor is in the range from 0.15 to 3.84 with a median of 1.20 for BL Lacs, and from 0.31 to 7.96 with a median of 2.03 for FSRQs, which indicates that FSRQs are stronger Doppler boosted than BL Lacs objects. This conclusion is consistent with estimations from other literatures \citep{Ghi93, Hov09, Fan13, Lio18, Che18}. We believe that FSRQs may have a smaller viewing angle $\theta$ relative to BL Lacs, hence resulting in a larger $\delta$ since $\delta=[\Gamma(1-\beta\cos\theta)]^{-1}$.

For confirming the reliability of our estimated outcomes, we once again cross-checked our sample with \citet{Fan09RAA}. A tight correlation that $\delta_{\rm TW}=(0.85\pm0.15)\delta_{\rm F09}+(1.02\pm0.26)$ with $r=0.61$ and $P=7.96\times10^{-7}$ for 54 sources in common (16 BL Lacs+38 FSRQs) has been reported.       

\begin{figure}
   \centering
   \includegraphics[width=12cm]{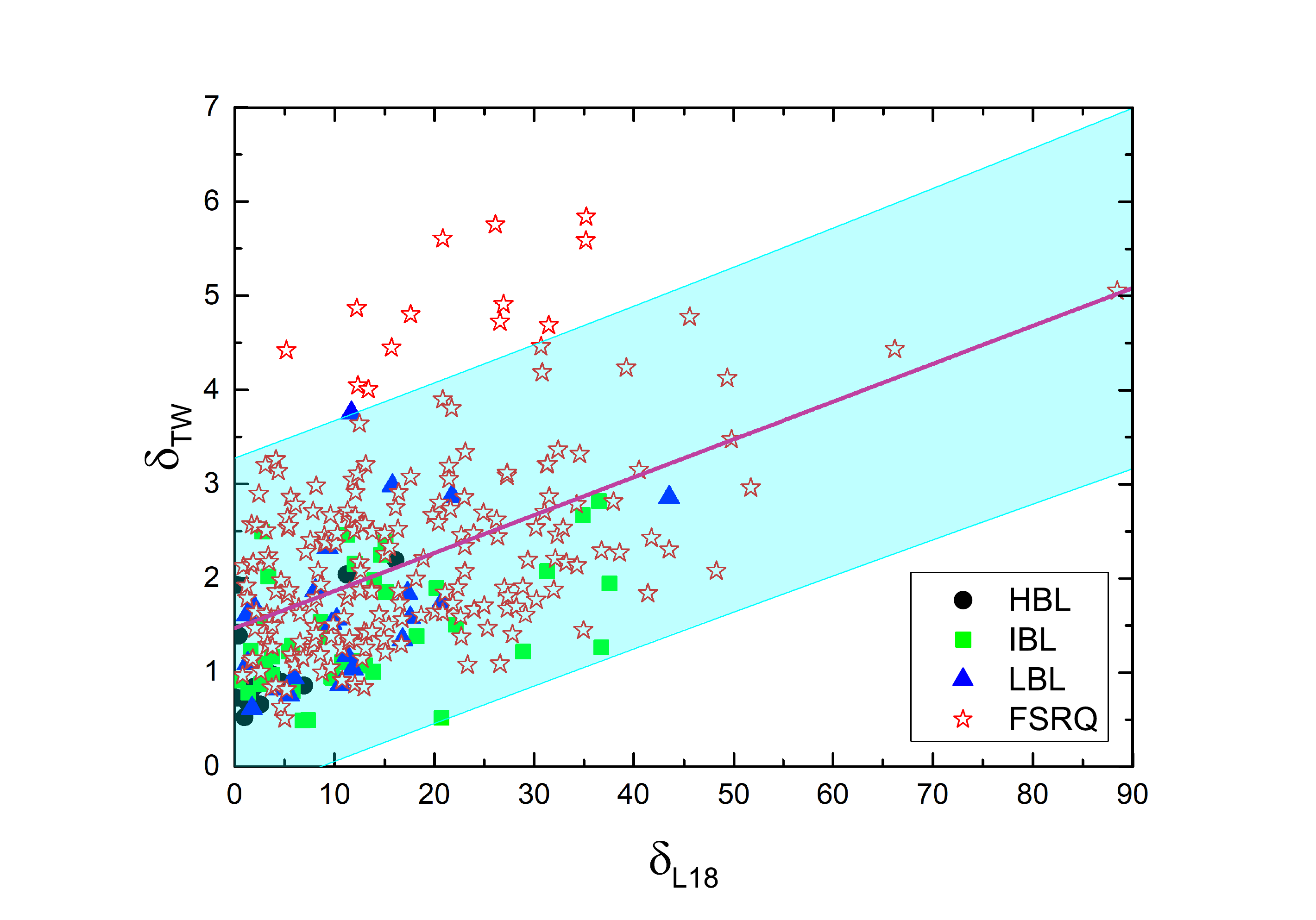}
   \caption{Plot of the correlation between the estimated Doppler factor $\delta$ derived from this paper (denoted by a subscript `TW') and from \citet{Lio18} (denoted by a subscript `L18'). The cyan shadowed area refers to the prediction band at $95\%$ level with respect to the best-fitting labelled by a {\it pink solid} line.}
     \label{fig11}
\end{figure}

Utilising the radio light curves modeling as a series of flares characterized by an exponential rise and decay, \citet{Lio18} estimated the variability Doppler factor ($\delta_{\rm var}$) for a larger sample comprised of 837 blazars including 167 BL Lacs and 670 FSRQs. 
After cross-checking with our sample, 282 sources in common are compiled, which contains 13 HBLs, 40 IBLs, 22 LBLs and 207 FSRQs. We look for the correlation and find $\delta_{\rm TW}=(0.04\pm0.004)\delta_{\rm L18}+(1.47\pm0.09)$ with $r=0.49$ and $P\sim0$ (Figure \ref{fig11}). There are two FSRQs reported extremely large Doppler factor in \citet{Lio18}, TXS 0446+112 (i.e., 4FGL J0449.1+1121) with $\delta_{\rm var}=88.44$ and S5 0212+73 (i.e., 4FGL J0217.4+7352) with $\delta_{\rm var}=66.21$. Our estimations also give comparatively high $\delta$ on these two sources, 5.05 and 4.43 for the former and the latter one, respectively. The correlation is still significant with $r=0.45$ and $P=3.77\times10^{-15}$ when these two point are excluded. This tight relation between Doppler factor derived from the radio variability and this work connotes that (i) our estimated results on Doppler factor are reliable; (ii) since the $\delta$ deducted from this present paper is established on the $\gamma$-ray behaviour and also X-ray behaviour, we provide an evidential hint that there is association between radio emission and $\gamma$-ray emission, suggesting that the $\gamma$-ray and radio regions possibly share the same relativistic effects, and the SSC mechanism may be responsible for the emission from radio to $\gamma$-rays for BL Lacs objects, while EC mechanism is perhaps answerable for FSRQs.            

\citet{Pei20PASA} presented an effective approach to estimate the $\gamma$-ray Doppler factor ($\delta_{\gamma}$),
\begin{equation}
\delta_{\gamma}\ge\left[{1.54\times10^{-3}\displaystyle\left(1+z\right)^{4+2\alpha}\left(\frac{d_{L}}{\rm Mpc}\right)^{2}\left(\frac{\Delta T}{\rm h}\right)^{-1}}\left(\frac{F_{\rm 1\,keV}}{\mu \rm Jy}\right){\left(\frac{E_{\gamma}}{\rm GeV}\right)^{\alpha}}\right]^{\frac{1}{4+2\alpha}}
\label{eq23}
\end{equation} \citep[see also][]{Fan13RAA, Fan14},
where $\alpha$ is the X-ray spectral index ($F_{\nu X}\propto\nu_{X}^{-\alpha}$), $h_{75}=$H$_{0}$/75, $\Delta T_{5}=\Delta T/(10^{5})$s, $\Delta T$ is the timescale in units of hour ($\Delta T=1$ day was adopted), $F_{\rm 1\,keV}$ denotes the flux density at 1 keV in units of $\mu$Jy and $E_{\gamma}$ stands for the $\gamma$-ray photon energy in units of GeV. 

\begin{figure}
   \centering
   \includegraphics[width=12cm]{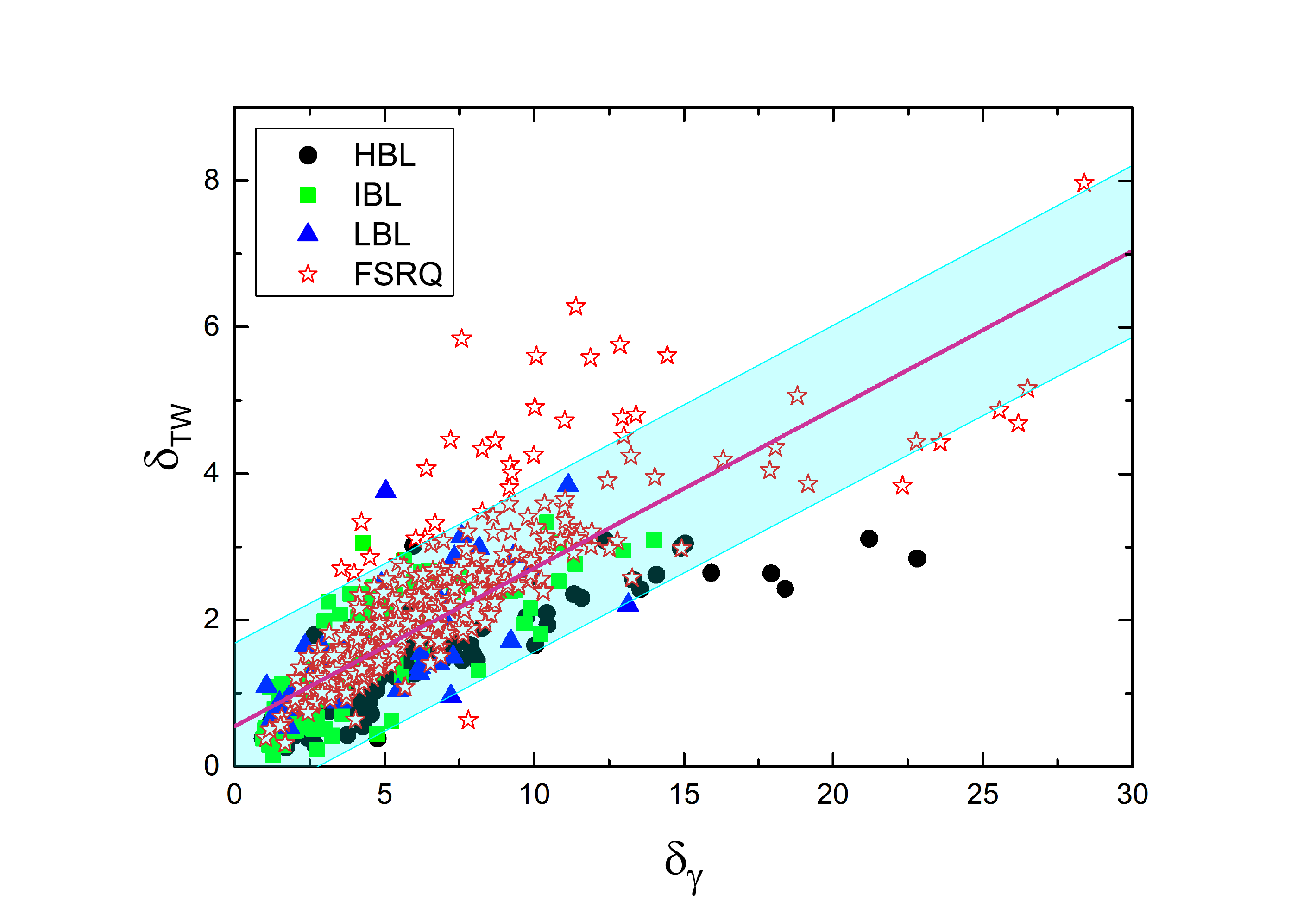}
   \caption{Plot of the correlation between the estimated Doppler factor $\delta_{\rm TW}$ derived from this paper and the $\gamma$-ray Doppler factor $\delta_{\gamma}$ from \citet{Pei20PASA}. The {\it pink solid} line signifies the best-fitting and the cyan shadowed area refers to the prediction band at $95\%$ level.}
     \label{fig12}
\end{figure}

In reference to the identical 809 sources, we obtained the $\gamma$-ray Doppler factor for FSRQs is higher on average than that for BL Lacs, namely $\langle\delta_{\gamma}\rangle_{\rm FSRQ}\simeq6.87$ and $\langle\delta_{\gamma}\rangle_{\rm BL\,Lac}\simeq4.31$, which suggests that the $\gamma$-ray emission of blazars is strongly beamed. We make the scatter plot of the Doppler factor derived from this work against the $\gamma$-ray Doppler factor in Figure \ref{fig12}. A clear tendency that $\delta_{\rm TW}$ increases with increasingly $\delta_{\gamma}$ is affirmed. The best-fitting is $\delta_{\rm TW}=(0.22\pm0.01)\delta_{\gamma}+(0.55\pm0.03)$ with $r=0.81$ and $P\sim0$ for the total sample.      

We need to remark that, the Doppler factors derived here are smaller than those obtained by others, and there are almost a quarter of sources with $\delta<1$. However, the lower than unity Doppler factor do not conflict with the beaming argument because we assume the emission to be originated from a cone with half angle $\Phi$ while others assume that the emission is isotropic. The difference between two circumstances lies on an enlarged factor of $\left(\frac{2}{1-\cos\Phi}\right)^{1/(4+\alpha_{\gamma})}$ according to Equation (\ref{eqA10}) (see Appendix \ref{App1}), i.e., $\delta_{4\pi}=\left(\frac{2}{1-\cos\Phi}\right)^{1/(4+\alpha_{\gamma})}\delta_{2\pi(1-\cos\Phi)}$ \citep{Che99, Fan05AA}. Thus one can obtain the corrected values of Doppler factor ranging from 0.70 to 17.04 with a median of 3.52 for our sample ($\delta$ for only 3 sources are less than unity in this case).     

Blazars jets are known to show fast variability, boosted emission, and apparent superluminal motion of jet components. These extreme properties are believed to be associated with the relativistic beaming effect dominating the emission from the jet, which can be quantified by the Doppler factor $\delta$. Unfortunately, up to now there is no direct method to measure either $\theta$ or $\Gamma$. Thus many subsidiary methods have been proposed for estimating the Doppler factor since $\delta$ is one of the most important parameters in the blazar paradigm and leading us to understand the energetics at large scales of their jets. In this paper, we work on the calculation of $\delta$ derived from our model for 809 {\it Fermi}-detected blazars. We find our results to be smaller compared with others since we consider the emission is non-isotropic that the $\gamma$-rays are from a cone with solid angle $\Omega=2\pi(1-\cos\Phi)$ whereas others assume that the emission is isotropic, in other words, $\Omega=4\pi$. We end up the discussion here. More detailed interpretation of Doppler effect and $\delta$ can be found in our previous papers \citep[e.g.,][]{Fan09, Fan13, Pei20PASA, Zha20, Ye21}.


\subsection{Propagation angle, $\Phi$}

Generally, the observed luminosity is calculated by assuming the emission is isotropic. However, many observed properties in some $\gamma$-ray loud blazars such as high luminosity, rapid variability and superluminal motion suggests that the $\gamma$-ray emission is strongly beamed. Henceforth, starting from the arguments by \citet{BK95}, the phenomenon that only the $\gamma$-rays within the propagation angle are visible, i.e. $\tau\le1.0$, has been discussed by many authors \citep[e.g.,][]{Che99, Fan05AA, Fan09RAA}. Thus, we can assume that the beamed $\gamma$-ray emission arises from a certain solid angle $\Omega=2\pi(1-\cos\Phi)$. In this present paper, our calculations show that the values of $\Phi$ for BL Lacs are in the range from 3.84$^\circ$ to 83.97$^\circ$ with a median of 16.33$^\circ$ , and from 7.33$^\circ$ to 84.31$^\circ$ with a median of 18.08$^\circ$ for FSRQs. Here we also cross-check our sample with \citet{Fan09RAA} for verifying the consistency. They obtained the average values of $\Phi$ for BL Lacs and FSRQs is 29.72$^\circ$ and 28.17$^\circ$, respectively, in a wide range from 3.02$^\circ$ to 83.31$^\circ$. Our derived results are well correlated with \citet{Fan09RAA} with the correlative coefficient $r=0.60$ and a chance probability of $P<10^{-6}$. \citet{Che99} obtained 13.0$^\circ$ to 39.2$^\circ$ with the average value of 24.6$^\circ$ for those 7 selected $\gamma$-ray loud blazars. \citet{Fan05AA} also reported their derived $\Phi$ is in the range from 8.91$^\circ$ to 56.49$^\circ$. Therefore, our derived values are consistent with other authors.     

\begin{figure}
   \centering
   \includegraphics[width=8.96cm]{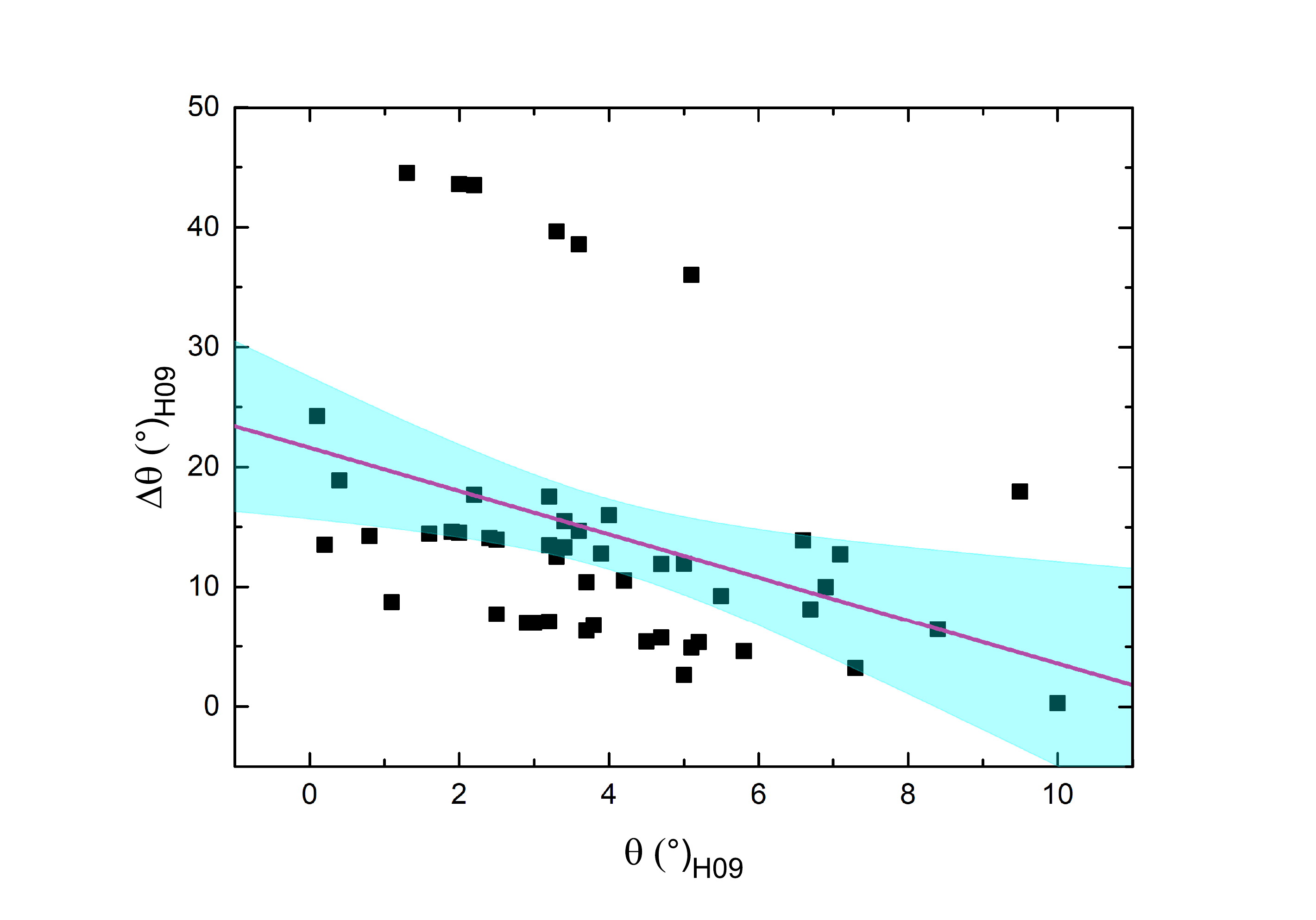}
   \includegraphics[width=8.96cm]{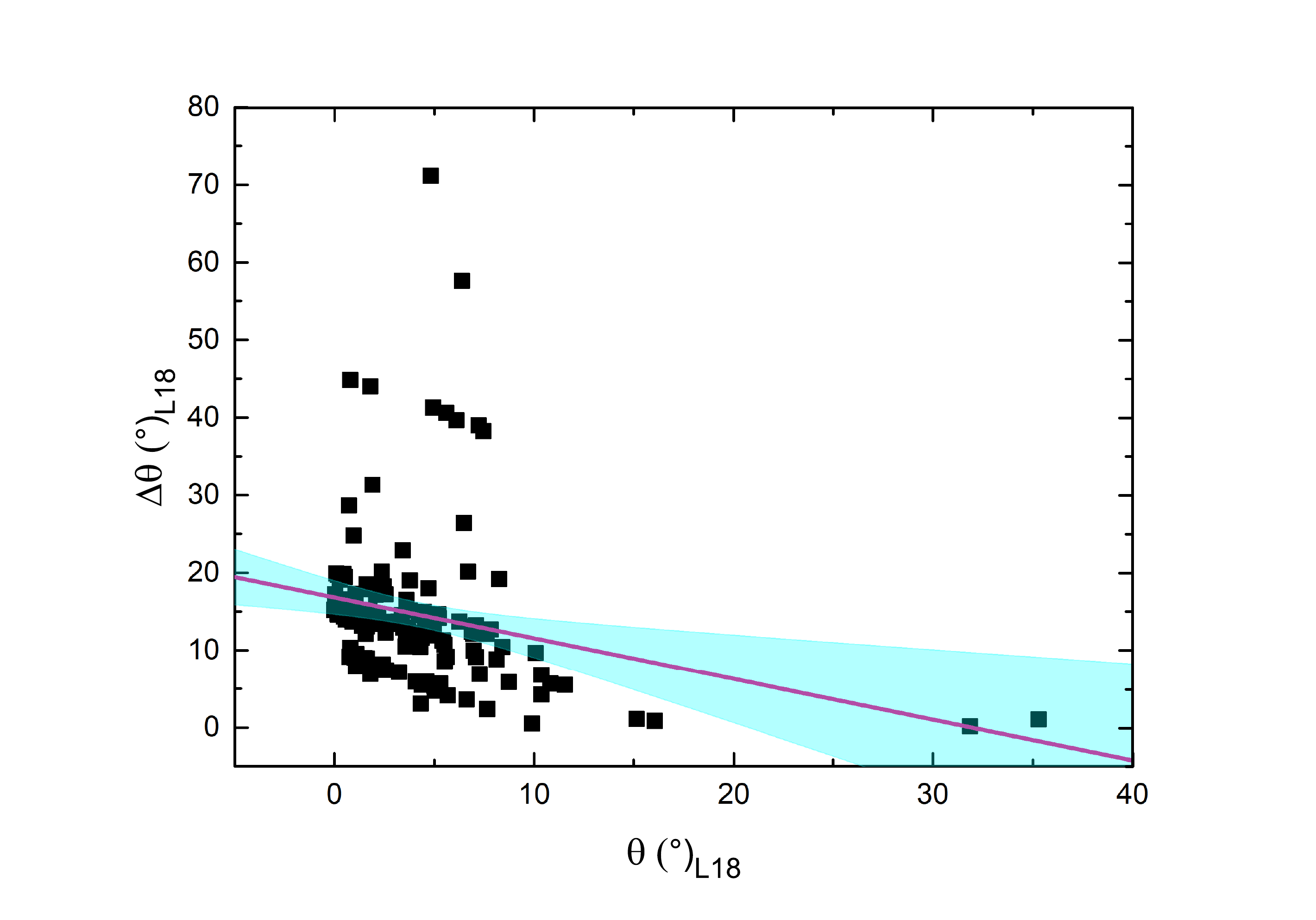}
    \caption{Correlation between $\Delta\theta=\Phi-\theta$ and $\theta$ for the cross-checked sample with \citet{Hov09} (left panel) and \citet{Lio18} (right panel). The {\it pink solid} lines refer to the best-fittings and the cyan shaded areas signify the confident level at 95$\%$.}
    \label{fig14}
\end{figure}

In the isotropic emission case, the $\gamma$-rays can be detected at any angle, however, in the scenario of non-isotropic emission, the emission is produced in a cone of a solid angle of $\Omega$, which yields the $\gamma$-rays would not be seen at any angle. Blazars are a subclass of AGNs, having their ultra-relativistic jets closely aligned to the line of site of an observer on Earth. Their small viewing angles result in the strong beaming effect, which can explain most of the physical properties of blazars. Since emitting high $\gamma$-ray radiation is a typical characteristic lying in the blazars, thus the angle we can detect the $\gamma$-rays should be greater than the viewing angle between the jet and the line-of-sight, i.e. $\Phi\ge\theta$. For probing this relation, we cross-checked our sample with \citet{Hov09} and \citet{Lio18}, respectively. \citet{Hov09} had calculated the variability Doppler factors for 87 sources by using the observations data at 22 and 37 GHz and from Very Long Baseline Interferometry (VLBI). Using apparent jet speed data, 62 blazars were given the Lorentz factors and viewing angles. They found almost all the sources in their sample are seen in a small viewing angle of less than 20$^\circ$, and FSRQs have a smaller $\theta$ than BL Lacs do. There are 51 sources in common after cross-check. We found their derived  propagation angles $\Phi$ are all larger than their viewing angle $\theta$, except for one source 4FGL J1806.8+6949 since \citet{Hov09} reported a quite large value of $\theta=57.3^\circ$ and our result shows $\theta=10.37^\circ$. We exclude this source in the following discussion. Firstly we cannot achieve a well correlation between $\Phi$ and $\theta$ for these 50 sources. Whereas, we obtain an interesting finding that the difference between $\Phi$ and $\theta$ decreases with increasing $\theta$. The left panel in Figure \ref{fig14} has shown this well correlation as $\Delta\theta=-(1.80\pm0.66)\theta+(21.59\pm2.95)$ with correlation coefficient $r=-0.37$ and a chance probability of $P=0.008$. Here, we denote $\Delta\theta=\Phi-\theta$.    

Similarly, \citet{Lio18} has estimated the viewing angles for 238 sources, 160 of which have been detected by {\it Fermi}-LAT. They found non-{\it Fermi}-detected sources have on average larger viewing angles than {\it Fermi}-detected ones. We cross-checked our sample with \citet{Lio18} and 152 blazars are in common. However, four sources are excluded because their $\theta$ are smaller than $\Phi$. They are 4FGL J1015.0+4926, J1058.6+5627, J2055.5+7752 and J2148.6+0652. Again, we do not find the correlation between their $\gamma$-ray propagation angles and viewing angles, but the anti-correlation between $\Delta\theta$ and $\theta$ are also discovered (see the right panel in Figure \ref{fig14}). The linear regression shows $\Delta\theta=-(0.53\pm0.18)\theta+(16.81\pm1.09)$ with correlation coefficient $r=-0.24$ and a chance probability of $P=0.003$.          
                
This outcome implies that the larger viewing angle, the closer for whom to approach to $\gamma$-ray propagation angle. We believe that this can be explained by the $\gamma$-rays are assumed to originate from a cone with a solid angle of $\Omega$. When we observe blazars, the $\gamma$-ray emission should be detected in the meantime. Based on the unified model of AGN \citep{UP95}, when the viewing angle is becoming larger, we would observe radio galaxies, e.g. FRIs and FRIIs. Some of them are showing $\gamma$-ray emission, indicating that the $\gamma$-ray propagation angles for these radio galaxies are much larger compared to blazars. Therefore, considering the constraint that blazars are observed, the viewing angle $\theta$ would be closer to the $\gamma$-ray propagation angle $\Phi$ with increasing $\theta$.

Finally, from the distribution of propagation angles, we can find that $90\%$ of BL Lacs and $83\%$ of FSRQs are located in the 1$\sigma$ confidence intervals with respect to their medians of $16.33^\circ$ and $18.08^\circ$, i.e. $\Phi=3.87^\circ-28.79^\circ$ for BL Lacs and $\Phi=10.00^\circ-26.16^\circ$ for FSRQs, separately, illustrating that the propagation of $\gamma$-rays forms a cone with respect to the axis of the accretion disk. \citet{MR94} also shows that the X-ray cone with propagation angle of $\Phi=15^\circ-40^\circ$ for BL Lacs. Our results are consistent with their conclusion.

\subsection{The $\gamma$-ray Emission Region}

The location of $\gamma$-ray-emitting in blazars is still an unresolved and opening problem. Constraining the production site of $\gamma$-ray emission can help us to comprehend the jet physics in blazars. This location implies the region where the bulk energy of the jet is converted to an energy distribution fo high-energy particles, and also determines the radiative cooling processes in leptonic and hadronic emission models. Aiming to this problem, many methods have been proposed \citep[e.g.][]{Jor01(a),Jor01(b), Jor10, Tav10, Tav13, Agu11(a), Agu11(b), Dot12, Yan12, Yan18, BE16, Wu18}. Generally, two scenario arise---the near site and the far site of the $\gamma$-ray regions. In the near site scenario, the electron energy is believed to be dissipated within the BLR \citep[e.g.][]{GM96, Geo01}, which locates at a distance of $<0.1-1$ pc from the SMBH, whereas in far site scenario, the dissipation of electron can be several parsecs away from the central engine \citep[e.g.][]{Lin05,Mar08, Zhe17}, where the dusty molecular torus (MT) are obligated to the dominating population of the target photons.          

\citet{Zhe17} performed a model-dependent method to determine the production site of $\gamma$-ray region for 36 FSRQs and obtained that the emission region is located at the range from 0.1 to 10 pc, i.e. outside the BLR but within the MT, which supports the far site scenario. Based on the measurements of the core-shift effect, the relation between the magnetic field strength ($B'$) in the radio core of the jet and the dissipation distance ($R_{\rm diss}$) of these radio core from the central SMBH can be derived. \citet{Yan18} applied this method with the observations of a FSRQ PSK 1510-089 (=4FGL J1512.8-0906) and BL Lacertae (=TXS 2200+420 or 4FGL J2202.7+4216). They found $R_{\rm diss}<0.5$ pc for hadronic model and $R_{\rm diss}<3.5$ pc for leptonic model for PSK 1510-089, while for BL Lacertae, $R_{\rm diss}<0.01$ pc for hadronic model and $R_{\rm diss}<0.02$ pc for leptonic model were reported, respectively. \citet{Ach21} argued that the $\gamma$-ray emission region locates within both the BLR and the MT from investigations that temporal and spectral analysis of $\gamma$-ray flux from selected brightest {\it Fermi}-detected-FSRQs.          

From our model presented in Section \ref{sec3}, the location of $\gamma$-ray emission dissipation, $R_{\gamma}$, can be determined by solving the equation,
\begin{eqnarray}
&R_{\gamma}^{2}=R^{2}+d^{2}+\lambda^{2}+2\lambda(R\sin\Phi+d\cos\Phi), \nonumber \\
&\sin\omega=\displaystyle\frac{R+\lambda\sin\Phi}{R_{\gamma}},
\label{eq24}
\end{eqnarray} 
where $\omega=\kappa\Phi$. In our estimation, $\kappa=0.1$ and $R=10R_{g}$ are adopted. Substituting our derived results of the distance $d/R_{g}$ and propagation angle $\Phi$, we can obtain the location of $\gamma$-ray-emitting region $R_{\gamma}$ to the central SMBH.         

\begin{figure}
   \centering
   \includegraphics[width=12cm]{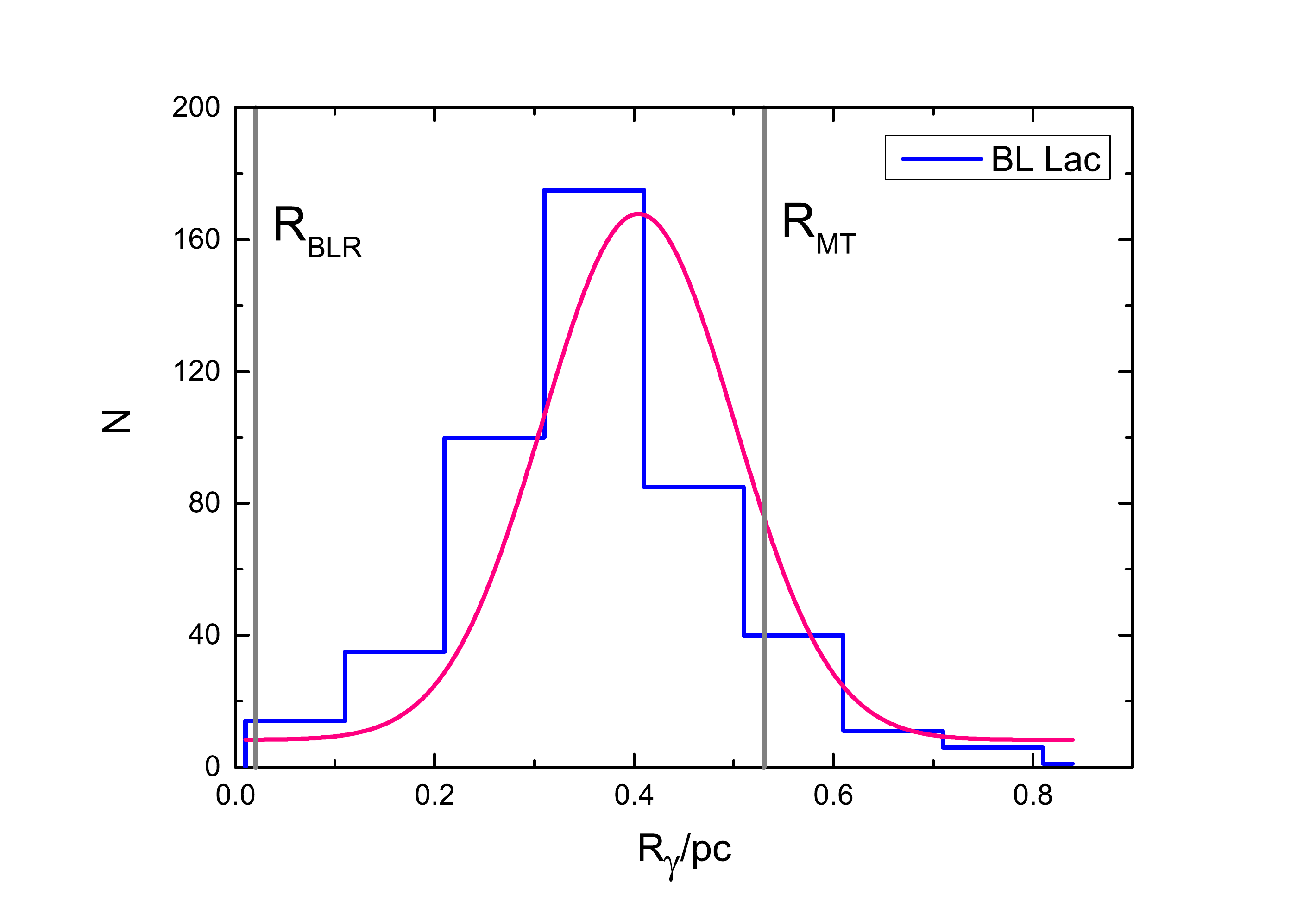}
   \includegraphics[width=12cm]{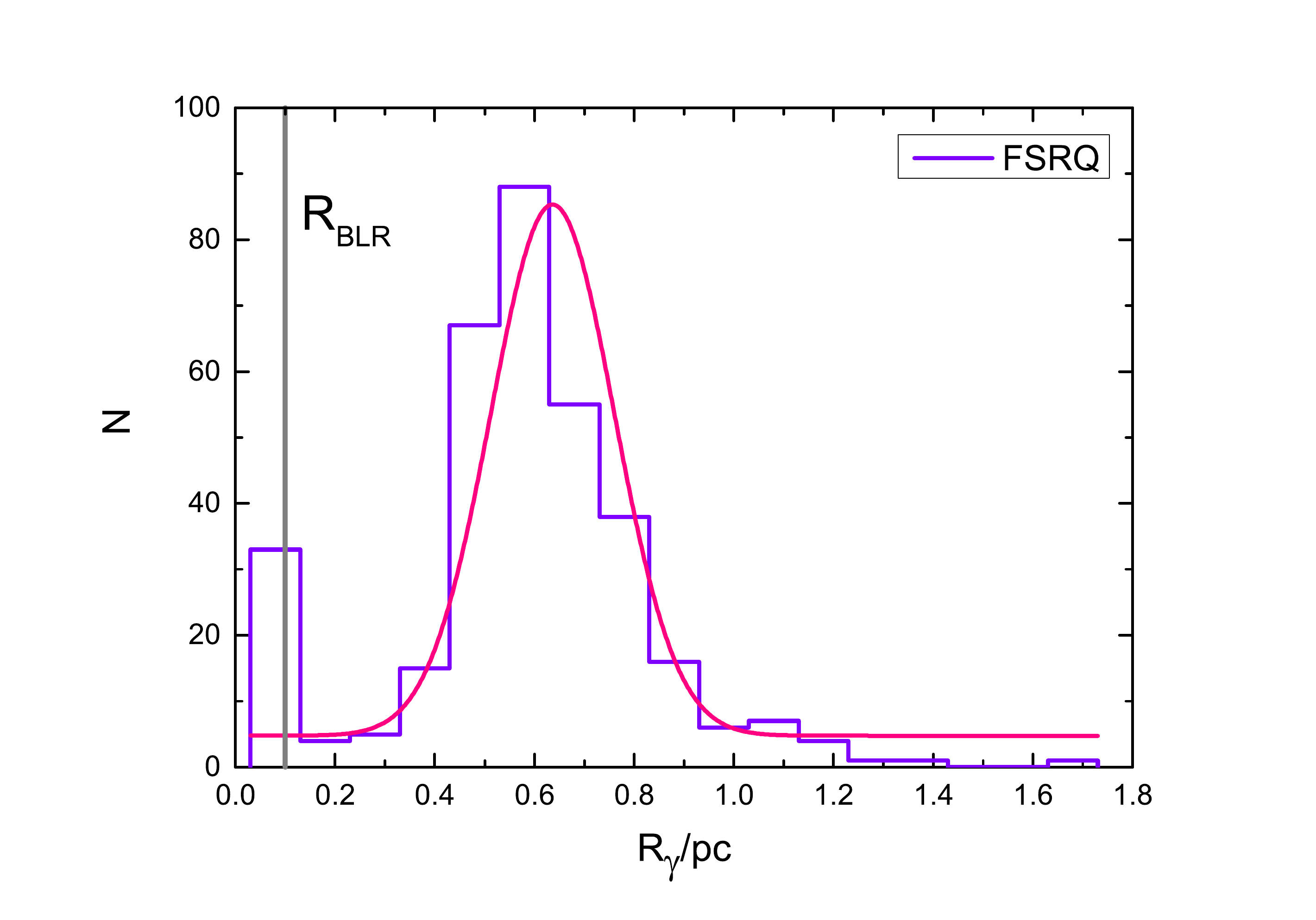}
    \caption{Distributions of the $\gamma$-ray emission location $R_{\gamma}$ estimated in this work.}
    \label{fig15}
\end{figure}

The histograms of $R_{\gamma}$ for 468 BL Lacs and 341 FSRQs are displayed in Figure \ref{fig15}, where $R_{\gamma}$ is in units of pc. The distribution of BL Lacs is in the range of 0.01 pc to 0.84 pc with a average of $0.36\pm0.13$ pc, while pervading a wider extent for FSRQs, spanning from 0.03 pc to 1.69 pc with a average of $0.58\pm0.25$ pc. We also perform the Gaussian fitting. Regarding BL Lacs $\mu=0.40\pm0.01$ and $\sigma=0.10\pm0.01$ with $P<10^{-6}$ are ascertained, whereas $\mu=0.64\pm0.01$ and $\sigma=0.12\pm0.02$ with $P<10^{-8}$ for FSRQs. Along with the probability $P=3.66\times10^{-63}$ from K-S test between two distributions, we can conclude that the $R_{\gamma}$ for FSRQs is significantly on average larger than that for BL Lacs.              

Furthermore, the average values of $R_{\gamma}$ for LBLs, IBLs and HBLs are $0.40\pm0.13$ pc, $0.36\pm0.14$ pc and $0.34\pm0.12$ pc, respectively. Thus LBLs occupy the right hand side of normal distribution, and are also closer to FSRQs' average. We consider this may be on account of the same evolution for LBLs and FSRQs. To verify it, we perform the K-S test and found the probability for they coming from the same parent population is $P=1.4\%$, which provides another evidence for the changing-look blazars.    

Lastly, we can see there is a distinct pile locating at the leftmost of FSRQs' distribution, which there are 35 out of overall 341 sources, for their site of $\gamma$-ray emission, are in the scope between 0.03 pc -- 0.14 pc. We consider these FSRQs having a comparatively small $R_{\gamma}$ are perhaps, again, changing-look sources that we mention above.    

The size of the BLR and dusty molecular torus (MT) can be estimated by means of the disk luminosity $L_{\rm Disk}$ \citep{Kas07, Ben09, GT09, Ghi14, Yan18}     
\begin{eqnarray}
&R_{\rm BLR}=10^{17}\left(\displaystyle\frac{L_{\rm Disk}}{10^{45}\,\rm erg\,s^{-1}}\right)^{1/2}\,\rm cm, \nonumber \\
&R_{\rm MT}=2.5\times10^{18}\left(\displaystyle\frac{L_{\rm Disk}}{10^{45}\,\rm erg\,s^{-1}}\right)^{1/2}\,\rm cm.
\label{eq25}
\end{eqnarray} 
Although we do not obtain $L_{\rm Disk}$ for each source in our sample, however, we can adopt 184 sources with available $L_{\rm BLR}$ looking for the divide between BL Lacs and FSRQs in Section \ref{sec5.1}, to estimate the sizes of BLR and MT. Mean values of $R_{\rm BLR}$ and $R_{\rm MT}$ for 37 BL Lacs are 0.02 pc and 0.53 pc, separately. On the other hand for 147 FSRQs, 0.1 pc and 2.54 pc for $R_{\rm BLR}$ and $R_{\rm MT}$ are acquired. Therefore, we use these values to constrain the locations of BLR and MT for our blazars.

We label the outer boundaries of BLR and dusty MT in grey in both panels of Figure \ref{fig15}. We found that, the $\gamma$-ray-emitting region for the vast majority of BL Lacs objects are beyond BLR except for two sources. $90.8\%$ of the sample (425 sources) are located outside the BLR and within the dusty MT, and closer to the MT than BLR from the Gaussian fitting. Similarly, most of the FSRQs also stay outside of the BLR and all of them are within the MT (we do not show this boundary in the figure since $R_{\rm MT}=2.53$ pc is rather far away from the whole distribution). Our finding that $90.9\%$ of the sources (310 out of 341) lie on the location between BLR and MT is in a good agreement with \citet{Zhe17}. We also obtain that, different from BL Lacs, the site of $\gamma$-ray-emitting region for FSRQs are much closer to the BLR boundary.

The GeV $\gamma$-ray emission in blazars is generally believed to be come from the IC process, in which the EC mechanism plays an important role in FSRQs and LBLs, while IBLs and HBLs normally can be explained by the SSC mechanism. In the EC process, the seed photons are determined by the production site of the $\gamma$-ray-emitting region, which may be dominated by an accretion disk, BLR, infrared torus, or cosmic background, corresponding to the $\gamma$-ray region located near the SMBH horizon, within the BLR, outside the BLR and within the MT, and much beyond the MT, respectively \citep{GT09}.

The BLR is a photon-rich environment, and the interaction between these photons and gamma-ray photons can result in photon-photon pair production. However, the MT has a much lower photon density than the BLR, indicating less likelihood of pair production in the MT than the BLR. Thus, the pair production reveals itself as an attenuation of the gamma-ray flux for emission emanating from the inner region of the BLR, while emission arising from the MT is not anticipated to have this spectral feature \citep{DP03, LB06, Ach21}.

Using the simultaneous or quasi-simultaneous multi-wavelength observations, \citet{Wu18} modeled the SEDs of 25 blazars by adopting a one-zone leptonic model, where the seed photons from the BLR and MT are considered in the EC process, and calculated the location of the $\gamma$-ray region for these blazars by means of assuming that the magnetic field strength derived in the SED fittings follows the magnetic field strength distribution as derived from the radio core-shift measurements. They also obtained that the emission emitting region may be outside the BLR at $R_{\gamma}\sim10R_{\rm BLR}$. Our present work differs from the generally popular methods that estimated the site of the $\gamma$-ray-emitting region, for instance, the SEDs modeling and variability timescales \citep[e.g.,][]{Der09, GT09, CW13, Kan14, Zhe17, Ach21}. We build up a photon-photon interaction model through the pair production process. Four fundamental physics parameters for $\gamma$-ray blazars can be constrained, including the distance along the axis to the site of the $\gamma$-ray production ($d/R_{g}$) which can be transformed into the location of $\gamma$-ray-emitting region $R_{\gamma}$. We find that $R_{\gamma}$ for FSRQs is on average larger than that for BL Lacs. The distribution for BL Lacs is between 0.01 pc -- 0.84 pc and 0.03 pc -- 1.69 pc for FSRQs, also known as staying outside the BLR and beyond the dusty MT for some BL Lacs. We consider that when the $\gamma$-ray emission is produced outside the BLR, the IC scattering could take place at the Thomson regime, where the GeV variability caused by electron cooling is energy dependent and faster at higher energy, and the GeV spectrum would have the same spectral index as the optical-infrared spectrum \citep{CW13}.

To conclude this section, we perform an effective method based on the $\gamma-\gamma$ pair production to estimate the location of $\gamma$-ray emission region. The whole sample is in the range from 0.01 pc to 1.69 pc. \citet{Ghi13} pointed out the most efficient location to produce the largest amount of $\gamma$-rays is at 0.1 pc to 1 pc, where there is the largest amount of seed photons at the maximum $\Gamma_{\rm var}$.

\subsection{Other related-parameters estimation}       

Above we have discussed four parameters obtained from this work. In this last subsection, we intend to estimate other related-parameters which are possible to be further deduced by means of these four parameters, for instance, the black hole mass $M$, the most important parameter we derived. 

The spin of SMBHs is coming into our sight in the first place since the power of relativistic jets of AGNs depends on the spin and the mass of the central SMBHs, as well as the accretion. The spin can be described by a dimensionless parameter $j$, defined as $j\equiv Jc/(GM^{2})$, where $J$ is the spin angular momentum of the black hole. Note that, $j$ is sometimes expressed in the symbol $a$ or $a_{\ast}$ in other work. We can calculate $j$ by using the following equation \citep{Dal19, Che21},
\begin{equation}
\displaystyle j=\frac{2\sqrt{f(j)/f_{\rm max}}}{f(j)/f_{\rm max}+1},
\label{eq26}
\end{equation}
where $f(j)$ is the spin function, which can be determined from 
\begin{equation}
\displaystyle \frac{f(j)}{f_{\rm max}}=\left(\frac{L_{j}}{g_{j}L_{\rm Edd}}\right)\left(\frac{L_{\rm bol}}{g_{\rm bol}L_{\rm Edd}}\right)^{-0.43}.
\label{eq27}
\end{equation}
Here $f(j)$ is normalized by its maximum value $f_{\rm max}$, that is, the value of $f(j)$ when $j=1$. In our calculation, $g_{j}=0.1$ and $g_{\rm bol}=1$ are used \citep{Dal19}. $L_{j}$ is the beam power of jets, which was originally estimated via the radio luminosity at 151 MHz \citep{Wil99}. However, an empirical relationship between the beam power and bolometric luminosity both are in units of Eddington luminosity is found by previous work \citep[e.g.][]{MH07, Fos11, Dal18, Pio20},
\begin{equation}
\displaystyle \log\frac{L_{j}}{L_{\rm Edd}}=\alpha\log\frac{L_{\rm bol}}{L_{\rm Edd}}+\beta,
\label{eq28}
\end{equation}  
where $\alpha$ and $\beta$ are best-fitting constants. We take $\alpha=0.41\pm0.04$ and $\beta=-1.34\pm0.14$ from \citet{Dal18}.     

We collect the bolometric luminosity from \citet{Fan16}, and 425 BL Lacs and 297 FSRQs comprise our sub-sample after cross-checked, then we estimate the spin of black holes for these sources. Using Equation (\ref{eq28}), (\ref{eq27}) and (\ref{eq26}), we obtain that the average spins for BL Lacs and FSRQs are $\left<j\right>_{\rm BL\,Lac}=0.51\pm0.20$ and $\left<j\right>_{\rm FSRQ}=0.55\pm0.20$, respectively. The K-S test yields the significance level probability for the null hypothesis that BL Lacs and FSRQs are drawn from the same distribution $P=0.001$ and the statistic $d_{\rm max}=0.14$. Given that FSRQs may have stronger accretion disk compared to BL Lacs which we have discussed in Section \ref{sec5.1}, an indicative conclusion that FSQRs perhaps have more prominent outflow effect within the black hole system than that of BL Lacs is reached. Our findings also suggest that the spin of SMBHs and accretion can power the relativistic jets.        

The magnetic fields play a critical role in jet formation and accretion disk physics \citep{BZ77, Zam14, Bla19}.  Together with the black hole mass, the spin of the black hole and the magnetic field strength ($B$), these three parameters couple the jet power. An accretion disk can be formed through matter falling onto the black hole, and the angular momentum can be lost by way of viscosity or turbulence \citep{Ree84} or magnetic field processes \citep[e.g.,][]{CS13} or via outflow.

The total magnetic field strength of the accretion disk can be estimated using \citep[e.g.,][]{Dal19},     
\begin{equation}
\displaystyle\left(\frac{B}{10^{4}\,\rm Gs}\right)=3.16\left(\frac{L_{\rm bol}}{g_{\rm bol}L_{\rm Edd}}\right)^{0.215}\left(\frac{\kappa^{2}_{B}}{M_{7}}\right)^{1/2},
\label{eq29}
\end{equation}  
and we adopt $\kappa_{B}=6$ \citep{Ree84} and $g_{\rm bol}=1$. Then we can obtain $B$ in units of Gs when substituting the derived $M_{7}$, $\left<\log B\right>_{\rm BL\,Lac}=4.54\pm0.26$ and $\left<\log B\right>_{\rm FSRQ}=4.75\pm0.32$ are ascertained, respectively. The K-S test shows that $d_{\rm max}=0.26$ with $P$-value of $9.05\times10^{-11}$. Our result on the accretion disk magnetic field strength of $\gamma$-ray blazars is in consonance with reporting by other authors \citep[e.g.,][]{Gar10, Mik15, Pio15, Che21}.   

The injected $\gamma$-ray compactness can be defined by 
\begin{equation}
\ell_{\gamma}=\displaystyle\frac{L_{\gamma}\sigma_{\rm T}}{4\pi R_{\gamma}m_{\rm e}c^{3}},
\label{eq10}
\end{equation}  
where $\sigma_{\rm T}$ is the Thomson cross section. Using $\gamma$-ray-emitting location $R_{\gamma}$ derived in this work, we can ascertain the $\gamma$-ray compactness, having the average value of $\left<\ell_{\gamma}\right>_{\rm BL\,Lac}=-2.34\pm0.79$ and $\left<\ell_{\gamma}\right>_{\rm FSRQ}=-1.47\pm0.38$, respectively. 

This parameter can be indicative for several interesting implications of photon quenching on compact $\gamma$-ray sources and emission models of $\gamma$-rays, providing the possibility that high-energy photons can pair-produce on soft target photons instead of escape in compact sources \citep[e.g.,][]{Jel66}, which suggests that the photon-photon annihilation could be not only as quenching of $\gamma$-rays, but also as sources of electron-positron pairs inside non-thermal compact sources \citep[e.g.,][]{Gui83, ZL85, Sve87}. \citet{PM11} also pointed out that the $\gamma$-rays would escape without any attenuation in one crossing time if there is no any substantial soft photon population within the source.    

To sum up, although the above three parameters cannot be derived directly from our model presented in this work, we are still able to estimate them by means of our investigated results in this paper.

\section{SUMMARY} \label{sec6}

In this paper, the optical depth of a $\gamma$-ray traveling in the field of a two-temperature disk and the beaming effect have been used to determine four fundamental physics parameters depicting the framework of $\gamma$-ray blazars, which include the upper limit of central black hole mass $M$, the Doppler factor $\delta$, the distance along the axis to the site of the $\gamma$-ray production $d$ (which can be transformed into the location of $\gamma$-ray-emitting region $R_{\gamma}$) and the propagation angle with respect to the axis of the accretion disk $\Phi$. Following in the footsteps of \citet{BK95}, we employ the same method firstly proposed from there and a sample of 809 {\it Fermi}-LAT-detected blazars has been compiled to derive $M$, $\delta$, $R_{\gamma}$ and $\Phi$. Only one source, 3C 279, had been discussed in \citet{BK95}, we enlarge the $\gamma$-ray blazars sample in this work and obtained that our estimations of $M$ and $\delta$ are consistency with other determinations. Besides, on the basis of \citet{BK95}, we bring forth several updated perspectives on revisiting the physical framework of blazars, for instance, putting forward the new divide between BL Lacs and FSRQs according to the accretion power in the disk, proposing an underlying transition field, exploring the relationship between the viewing angle and $\gamma$-ray propagation angle, and determining the $\gamma$-ray-emitting region. We draw our main conclusions as follows:

\begin{itemize}
  \item The black hole mass, along with mass accretion rate, is a fundamental property of blazars. We obtain medians of $M_{7}$ are 13.34 for BL Lacs and 16.24 for FSRQs ($M_{7}=10^{7}M_{\sun}$). Compared to other estimated diagnoses, we find our results on black hole mass are in a better agreement with deriving from the BLR property. 
   
  \item We put forward an updated demarcation between BL Lacs and FSRQs based on the relation between BLR luminosity and disk luminosity both measured in Eddington units first proposed by \citet{Ghi11}, that is, $L_{\rm disk}/L_{\rm Edd}=4.68\times10^{-3}$, indicating that there are some differences between BL Lacs and FSRQs on the accretion power in the disk. This dividing line on the ratio $L_{\rm Disk}/L_{\rm Edd}$ between BL Lacs and FSRQs also confirms the idea that the blazars' divide occur for the alteration in accretion regime.
  
   \item We propose a so-called `appareling zone' in the range from $L_{\rm BLR}/L_{\rm Edd}\simeq2.00\times10^{-4}$ to $8.51\times10^{-3}$, which stands for a potential transition field between BL Lacs and FSRQs that changing-look blazars may reside. We found five confirmed changing-look sources in our sample that are lying on this zone. They are 4FGL J10001.1+2911, J1043.2+2408, J1153.4+4931, J1422.3+3223 and J1503.5+4759, respectively.   
      
   \item The Doppler factor is a crucial parameter for discussing the beaming effect. We derive $\delta$ has a median of 1.20 for BL Lacs while of 2.03 for FSRQs. We ascertain relatively small values of Doppler factor due to our consideration that the emission is not isotropic, i.e., coming from a solid angle with $\Omega=2\pi(1-\cos\Phi)$.  
   
   \item We determine the location of $\gamma$-ray emission region, $R_{\gamma}$, which is principally constrained outside the BLR, and for some BL Lacs are also away from the MT. This supports the idea that the most efficient location to produce the largest amount of $\gamma$-rays is at 0.1 pc to 1 pc, where there is the largest amount of seed photons at the maximum $\Gamma_{\rm var}$.  
   
   \item We also estimate the spin of black hole, magnetic field strength and $\gamma$-ray compactness by dint of our derived results.

\end{itemize}

\acknowledgments

We are grateful to the anonymous referee for valuable comments and constructive suggestions, which help us to improve the manuscript. The work is partially supported by the National Natural Science Foundation of China (NSFC U2031201, NSFC 11733001, U2031112), Guangdong Major Project of Basic and Applied Basic Research (Grant No. 2019B030302001). Z.Y. Pei acknowledges support from National Science Foundation for Young Scientists of China (Grant 12103012). We also acknowledge the science research grants from the China Manned Space Project with NO. CMS-CSST-2021-A06, and the supports for Astrophysics  Key Subjects of Guangdong Province and Guangzhou City. The work is also supported by Guangzhou University (YM2020001). This research has made use of the NASA/IPAC Extragalactic Database (NED), which is funded by the National Aeronautics and Space Administration and operated by the California Institute of Technology. Part of this work is based on archival data, software or online services provided by the Space Science Data Center - ASI.

\appendix

\section{Model description and calculation process on the estimation of four fundamental physics parameters} \label{App1}

We give here the full description of the model used to deduct the equations for deriving these four parameters of blazars adopted in this work.

We consider a diagram of $\gamma$-ray propagation above a two-temperature disk surrounding the central black hole (see Figure \ref{fig1}). In this scenario, the strongly beamed $\gamma$-rays interact with the soft X-ray photons produced at all points on the disk within an angle between the $\gamma$-ray trajectory and the $z$-axis ($\Phi$). Since the optical depth $\tau$ is not greater than unity and based on the idea firstly proposed by \citep{BK95}, we can obtain an approximate empirical formula for optical depth building on a two-temperature disk scenario at an arbitrary angle of $\Phi$ \citep{Che99},
\begin{equation}
\tau_{\gamma\gamma}(M_{7},\Phi,d)=9\times\Phi^{2.5}\left(\displaystyle\frac{d}{R_{g}}\right)^{-\frac{2\alpha_{X}+3}{2}}+kM_{7}^{-1}\left(\frac{d}{R_{g}}\right)^{-2\alpha_{X}+3},
\label{eqA1}
\end{equation}
where $M_{7}$ denotes the black hole mass in units of $10^{7}M_{\sun}$, $\alpha_{X}$ is the X-ray spectral index and
\begin{equation}
k=4.61\times10^{9}\displaystyle\frac{\Psi(\alpha_{X})(1+z)^{3+\alpha_{X}}F'_{0}(1+z-\sqrt{1+z})^2}{(2\alpha_{X}+1)(2\alpha_{X}+3)}\times\left[\frac{\left(\frac{R_{o}}{R_{g}}\right)^{2\alpha_{X}+1}-\left(\frac{R_{\rm ms}}{R_{g}}\right)^{2\alpha_{X}+1}}{\left(\frac{R_{\rm ms}}{R_{g}}\right)^{-1}-\left(\frac{R_{0}}{R_{g}}\right)^{-1}}\right]\left(\frac{E_{\gamma}}{4m_{\rm e}c^{2}}\right)^{\alpha_{X}}.
\label{eqA2}
\end{equation}      
Here, $F'_{0}$ is the X-ray flux parameter in units of cm$^{-2}$s$^{-1}$, $R_{g}=\frac{GM}{c^{2}}$ is the Schwarzschild radius, $E_{\gamma}$ denotes the average energy of $\gamma$-ray emission, $R_{0}$ and $R_{\rm ms}$ refer to the inner and outer radii of the hot region of a two-temperature accretion disk \citep{BK95}, and $\Psi(\alpha_{X})$ is a function of the X-ray spectral index,
\begin{equation}
\Psi(\alpha_{X})\equiv\sigma_{\rm T}^{-1}\int_{0}^{1}2\rho(1-\rho^{2})^{\alpha_{X}-1}\sigma_{\gamma\gamma}(\rho)d\rho,
\label{eqA3}
\end{equation}
where $\rho$ is the velocity of the positron or electron in the center-of-momentum frame in units of $c$, $\sigma_{\rm T}$ is the Thomson cross section, and $\sigma_{\gamma\gamma}(\rho)$ signifies the exact cross section for $\gamma$-$\gamma$ pair production given by
\begin{equation}
\sigma_{\gamma\gamma}(\rho)=\displaystyle\frac{3}{16}\sigma_{\rm T}(1-\rho^{2})\left[(3-\rho^{4})\ln\left(\frac{1+\rho}{1-\rho}\right)+2\rho^{3}-4\rho\right],
\label{eqA4}
\end{equation}

The variability time scale ($\Delta T_{\rm D}$) can constrain the distance along the axis to the site of $\gamma$-ray emission region, which are expressed as
\begin{equation}
\displaystyle\frac{d}{R_{g}}=1730\times\frac{\Delta T_{\rm D}}{1+z}\delta M_{7}^{-1},
\label{eqA5}
\end{equation}
where $\frac{d}{R_{g}}$ denotes the distance in units of $R_{g}$, $\Delta T_{\rm D}$ is in units of days and $\delta$ is the Doppler factor defined as
\begin{equation}
\delta=\displaystyle\frac{1}{\Gamma(1-\beta\cos\Phi)},
\label{eqA6}
\end{equation}
$\Gamma$ is the Lorentz factor and $\beta$ is the bulk velocity in units of the speed of light $c$.

In the beaming model, the observed $\gamma$-ray luminosity can be expressed by the form of \citep{Fan05AA}
\begin{equation}
L_{\gamma}^{\rm obs}=\displaystyle\frac{\delta^{\alpha_{\gamma}+4}}{(1+z)^{\alpha_{\gamma}-1}}L_{\gamma}^{\rm in},
\label{eqA7}
\end{equation}
here $L_{\gamma}^{\rm in}$ is the $\gamma$-ray intrinsic luminosity in the comoving frame and $\alpha_{\gamma}$ is the $\gamma$-ray spectral index. Since the observed luminosity can be expressed as $L_{\gamma}^{\rm obs}=\Omega d_{\rm L}^{2}F_{\gamma}^{\rm obs}$, thus Equation (\ref{eqA7}) can be derived into
\begin{equation}
F_{\gamma}^{\rm obs}=(1+z)^{1-\alpha_{\gamma}}\delta^{\alpha_{\gamma}+4}L_{\gamma}^{\rm in}/\Omega d_{\rm L}^{2}.
\label{eqA8}
\end{equation}  
We can define an isotropy luminosity as $L_{\rm iso}=4\pi d_{\rm L}^{2}F_{\gamma}^{2}$, thus we can obtain
\begin{equation}
L_{\rm iso}^{45}=\displaystyle\frac{2.52\lambda\delta^{\alpha_{\gamma}+4}}{(1-\cos\Phi)(1+z)^{\alpha_{\gamma}-1}}M_{7},
\label{eqA9}
\end{equation}  
where we adopt $L_{\gamma}^{\rm in}=\lambda L_{\rm Edd}=\lambda1.26\times10^{45}M_{7}$, $\lambda$ is a parameter depending on specific $\gamma$-ray emission model and $L_{\rm iso}^{45}$ is in units of $10^{45}$ erg s$^{-1}$. Then the Doppler factor can be derived, i.e.
\begin{equation}
\delta=\left(\displaystyle\frac{L_{\rm iso}^{45}(1-\cos\Phi)(1+z)^{\alpha_{\gamma}-1}}{2.52\lambda{M_{7}}}\right)^{\frac{1}{{\alpha_{\gamma}+4}}}.
\label{eqA10}
\end{equation} 
When substituting Equation (\ref{eqA10}) into Equation (\ref{eqA6}), one can read
\begin{equation}
d(\Phi,M,L_{\rm iso})=AR_{g}(1-\cos\Phi)^{\frac{1}{{\alpha_{\gamma}+4}}},
\label{eqA11}
\end{equation} 
where 
\begin{equation}
A=1730\times\Delta T_{D}(1+z)^{-\frac{5}{\alpha_{\gamma}+5}}M_{7}^{-\frac{\alpha_{\gamma}+5}{\alpha_{\gamma}+4}}\left(\displaystyle\frac{L_{\rm iso}}{2.52\lambda}\right)^{\frac{1}{{\alpha_{\gamma}+4}}}.
\label{eqA12}
\end{equation} 

After substituting Equation (\ref{eqA11}) and Equation (\ref{eqA10}) into Equation (\ref{eqA1}), we ascertain
\begin{equation}
\tau(\Phi,M,L_{\rm iso})=\left[9\times\Phi^{2.5}(1-\cos\Phi)^{-\frac{2\alpha_{X}+3}{2\alpha_{\gamma}+8}}+kM_{7}^{-1}A^{-\frac{2\alpha_{X}+3}{2}}(1-\cos\Phi)^{-\frac{2\alpha_{X}+3}{\alpha_{\gamma}+4}}\right]A^{-\frac{2\alpha_{X}+3}{2}}.
\label{eqA13}
\end{equation}
Then we set $\tau(\Phi,M,L_{\rm iso})=1.0$ and thus $\frac{\partial\tau_{\gamma\gamma}}{\partial\Phi}|_{M}=0$, i.e.,
\begin{eqnarray}
\frac{\partial\tau_{\gamma\gamma}}{\partial\Phi}|_{M}&=&\left[22.5\times\Phi^{1.5}(1-\cos\Phi)-9\times\frac{2\alpha_{X}+3}{2\alpha_{\gamma}+8}\Phi^{2.5}\sin\Phi-\frac{2\alpha_{X}+3}{2\alpha_{\gamma}+4}kM_{7}^{-1}A^{-\frac{2\alpha_{X}+3}{2}}(1-\cos\Phi)^{-\frac{2\alpha_{X}+3}{2\alpha_{\gamma}+8}}\sin\Phi\right] \nonumber \\
&\times&\left[(1-\cos\Phi)^{-\frac{2\alpha_{X}+2\alpha_{\gamma}+11}{2\alpha_{\gamma}+8}}A^{-\frac{2\alpha_{X}+3}{2}}\right]=0,
\label{eqA14}
\end{eqnarray}
which yields
\begin{equation}
22.5\times\Phi^{1.5}(1-\cos\Phi)-9\times\frac{2\alpha_{X}+3}{2\alpha_{\gamma}+8}\Phi^{2.5}\sin\Phi-\frac{2\alpha_{X}+3}{2\alpha_{\gamma}+4}kM_{7}^{-1}A^{-\frac{2\alpha_{X}+3}{2}}(1-\cos\Phi)^{-\frac{2\alpha_{X}+3}{2\alpha_{\gamma}+8}}\sin\Phi=0.
\label{eqA15}
\end{equation}
Finally, we derive four equations \citep{Che99, Fan05AA, Fan09RAA}
\begin{eqnarray}
&\displaystyle\frac{d}{R_{g}}&=1730\times\frac{\Delta T_{\rm D}}{1+z}\delta M_{7}^{-1}, \nonumber \\
&L_{\rm iso}^{45}&=\displaystyle\frac{2.52\lambda\delta^{\alpha_{\gamma}+4}}{(1-\cos\Phi)(1+z)^{\alpha_{\gamma}-1}}M_{7}, \nonumber \\
&9\times&\Phi^{2.5}\left(\displaystyle\frac{d}{R_{g}}\right)^{-\frac{2\alpha_{X}+3}{2}}+kM_{7}^{-1}\left(\displaystyle\frac{d}{R_{g}}\right)^{-2\alpha_{X}-3}=1, \nonumber \\
&22.5&\times\Phi^{1.5}(1-\cos\Phi)-9\times\frac{2\alpha_{X}+3}{2\alpha_{\gamma}+8}\Phi^{2.5}\sin\Phi-\frac{2\alpha_{X}+3}{2\alpha_{\gamma}+4}kM_{7}^{-1}A^{-\frac{2\alpha_{X}+3}{2}}(1-\cos\Phi)^{-\frac{2\alpha_{X}+3}{2\alpha_{\gamma}+8}}\sin\Phi=0.
\label{eqA16}
\end{eqnarray}

\bibliography{Pei2021}{}
\bibliographystyle{aasjournal}

\end{document}